\documentclass[prb,showpacs,preprintnumbers,amsfonts,amssymb,amsmath,floats,twocolumn,aps]{revtex4}

\usepackage{graphicx}
\usepackage{dcolumn}
\usepackage{bm}
\usepackage[final,dvips]{epsfig}
\usepackage{bm}
\usepackage{color}

\newcommand{\be}{\begin{equation}}
\newcommand{\ee}{\end{equation}}
\newcommand{\ba}{\begin{eqnarray}}
\newcommand{\ea}{\end{eqnarray}}
\newcommand{\ff}[1]{{\bm #1}}

\newcommand{\Tr}{\mbox{Tr}}

\begin{document} 

\title{ 
Mott transition in one dimension: Benchmarking dynamical cluster approaches
} 

\author{Matthias Balzer and Werner Hanke}

\affiliation{
Institut f\"ur Theoretische Physik und Astrophysik, Universit\"at W\"urzburg, Germany
}

\author{Michael Potthoff}

\affiliation{
I. Institut f\"ur Theoretische Physik, Universit\"at Hamburg, Germany
}

\begin{abstract}
The variational cluster approach (VCA) is applied to the one-dimensional Hubbard model at zero
temperature using clusters (chains) of up to ten sites with full diagonalization and the Lanczos 
method as cluster solver.
Within the framework of the self-energy-functional theory (SFT), different cluster reference 
systems with and without bath degrees of freedom, in different topologies and with different 
sets of variational parameters are considered.
Static and one-particle dynamical quantities are calculated for half-filling as a function of $U$ 
as well as for fixed $U$ as a function of the chemical potential to study the interaction- and
filling-dependent metal-insulator (Mott) transition. 
The recently developed $Q$-matrix technique is used to compute the SFT grand potential.
For benchmarking purposes we compare the VCA results with exact results available from the Bethe
ansatz, with essentially exact dynamical DMRG data, with (cellular) dynamical mean-field theory
and full diagonalization of isolated Hubbard chains.
Several issues are discussed including convergence of the results with cluster size, the ability
of cluster approaches to access the critical regime of the Mott transition, efficiency in the 
optimization of correlated-site vs.\ bath-site parameters and of multi-dimensional parameter 
optimization. 
We also study the role of bath sites for the description of excitation properties and as charge 
reservoirs for the description of filling dependencies. 
The VCA turns out to be a computationally cheap method which is competitive with established 
cluster approaches.
\end{abstract} 

\pacs{
71.10.-w, 71.30.+h, 71.10.Fd
} 

\maketitle 

\section{Introduction}

One of the most fascinating effects of strong interactions among itinerant 
electrons is insulating behavior that is induced by electron correlations. 
\cite{Mot61,Mot90}
An understanding of the Mott insulating state and also of the transition 
from a correlated metal to a Mott insulator is important for many transition-metal 
oxides including the parent compounds of cuprate-based high-temperature 
superconductors. \cite{IFT98}
The Mott transition is generically described using lattice models with purely 
local interactions, such as the single-band Hubbard model. \cite{Hub63,Gut63,Kan63}
Correlations, however, are generally non-local or even long-ranged.
It is a highly non-trivial question to what extent the Mott transition is dominated 
by local correlations and whether or not local approximations, i.e.\ approximations 
neglecting non-local correlations, are able to capture the essence of the Mott 
physics. 

The implications of the dynamical mean-field theory, \cite{MV89,GKKR96,KV04} as a 
distinguished local approximation, have been worked out in detail in the past
and have been compared with numerous experiments on transition-metal oxides.
The DMFT is a mean-field approach in the sense that the effects of non-local 
spin or charge two-particle correlations on the one-particle excitation spectrum 
are neglected.
In case of layered, essentially two-dimensional compounds, for example, this is 
probably a too strong approximation.

Different generalizations of DMFT have been suggested in the past to overcome this 
limitation. \cite{MJPH05,TKH07,SJMD06,RKL06}
Cluster extensions \cite{MJPH05} which restore the effects of non-local correlations 
step by step with increasing cluster size, are conceptually simple and interesting 
approaches in this respect. 
The idea is that, besides the local (temporal) correlations, it is the effect of 
the {\em short-range} correlations, treated exactly in a cluster approach,
which dominates the physics of the Mott transition or at least the physics of the
Mott-insulating state.
As the importance of non-local correlations is expected to increase with 
decreasing lattice dimension, the most stringent test for a cluster approach
consists in its application to the one-dimensional lattice.
Ideally, starting from a (dynamical) mean-field picture of the transition, 
the decisive step forward should be done with the smallest clusters already, while a
further increase of the cluster size should add qualitatively unimportant corrections
only.
Since exact results from the Bethe ansatz are available for the Hubbard model 
in one dimension, \cite{LW68} this model can very well be used to test 
this idea and to benchmark different cluster approximations.

Apart from true extensions of DMFT, \cite{HTZ+98,LK00,KSPB01,OMMF03} where a
small cluster with $L_c$ correlated sites and a continuum ($n_s=\infty$) of 
uncorrelated (``bath'') sites is considered, there are also dynamical cluster 
approaches without any bath degrees of freedom (i.e.\ $n_s=1$), such as the 
cluster-perturbation theory (CPT) \cite{GV93,SPPL00} and the variational cluster 
approach (VCA). \cite{PAD03}
The self-energy-functional theory (SFT), \cite{Pot03a,PAD03} provides a unified 
framework for all these different cluster approximations which are characterized by 
different $L_c$ and $n_s$. 
Therefore, the purpose of this paper is to apply (within the SFT) different cluster 
approaches to the one-dimensional Hubbard model and to study the interaction- and the 
filling-controlled transition for benchmarking purposes. 

An obvious question concerns the rate of convergence to the exact solution 
which is approached in the $L_c=\infty$ limit. 
Depending on the cluster scheme used and on the quantity of interest, an exponential
or power-law dependence on $L_c$ can be inferred. \cite{MJ02a,BK02,AMJ05,BK05} 
These considerations, however, apply to the large-$L_c$ limit only while for 
practical purposes the quality of a given approximation for {\em small} $L_c$ 
is much more important and can only be estimated {\em a posteriori}.

It is also unclear whether or not the inclusion of bath sites makes a cluster 
approach more efficient, i.e.\ speeds up the convergence to the exact solution 
(see the discussion in Ref.\ \onlinecite{PAD03}).
Because of the exponential growth of the Hilbert space with increasing $n_s$,
this is of particular importance for methods that are based on a full (or Lanczos) 
diagonalization of the effective cluster model.

A simple and frequently used \cite{Dag94} cluster approximation is the ``direct'' 
cluster approximation where quantities for the infinite system are approximated 
by those of a finite cluster without any embedding of the cluster into a medium 
that mimics the disregarded cluster environment.
Due to particle-number conservation and due to the finite (and usually small) 
cluster size $L_c$, it is inconvenient to study filling dependencies using the
direct cluster approach.
The filling-controlled Mott transition, in particular, is hardly accessible in 
this way. 
Another interesting question is therefore, if and how a continuous filling 
dependence can be achieved by embedded, self-consistent or variational cluster
approaches. 

In the following Sec.\ \ref{sec:mott}, we briefly list some well-known properties 
of the Hubbard model in one dimension which are relevant for our study.
Sec.\ \ref{sec:cluster} provides a brief discussion of cluster approaches 
employing the exact diagonalization method at zero temperature. 
Some details of the SFT and of our numerical approach are given in Sec.\ \ref{sec:sft}.
The numerical results for the Mott insulator at half-filling and for the filling-controlled
Mott transition are presented and discussed in Secs.\ \ref{sec:half} and \ref{sec:fill},
respectively.
Sec.\ \ref{sec:sum} summarizes our main results.

\section{The Mott transition in the one-dimensional Hubbard model}
\label{sec:mott}

In one dimension, the single-band (grand-canonical) Hubbard model is given by 
\begin{equation}
  H 
  = 
  - t \sum_{i\sigma} \left( c^\dagger_{i\sigma} c_{i-1\sigma} + \mbox{H.c.} \right)
  - \mu \sum_{i\sigma} n_{i\sigma}
  + U \sum_{i\sigma} n_{i\uparrow} n_{i\downarrow} \: .
\end{equation}
Here, $c_{i\sigma}$ annihilates an electron at the site $i$ with spin projection
$\sigma=\uparrow,\downarrow$.
Furthermore, $n_{i\sigma}=c_{i\sigma}^\dagger c_{i\sigma}$ is the occupation-number
operator, the ground-state average of which is the site- and spin-independent filling
$n = \langle n_{i\sigma} \rangle$. 
We consider nearest-neighbor hopping only and set $t=1$ to fix the energy scale. 
Finally, $U$ denotes the strength of the local Coulomb repulsion, and $\mu$ is
the chemical potential.

The ground-state energy $E_0$ (of $H+\mu N$) can be calculated exactly \cite{LW68} 
by means of the Bethe ansatz.
For $\mu = U / 2$ the model is particle-hole symmetric and half-filled ($n=1$). 
If $L$ denotes the number of sites, we have
\begin{equation}
  E_0 / L = - 4 t \int_0^\infty dx \: \frac{J_0(x) J_1(x)}{x (1+\exp(xU/2t))} \: ,
\label{eq:gse}
\end{equation}
where $J_0,J_1$ are Bessel functions.
While the system is metallic for $U=0$, a Mott-insulating state is found for any
$U>0$ as can be seen from the exact expression \cite{Ovc69} for the single-particle gap:
\begin{equation}
  \Delta = \frac{16t^2}{U} \int_1^\infty dx \: \frac{\sqrt{x^2-1}}{\sinh(2\pi t x / U)} \: .
\label{eq:gap}
\end{equation}
The gap is finite for $U>0$ but exponentially small in the limit $U\to 0$, i.e.\ $\Delta \sim \exp(-1/U)$.
Opposed to the dynamical mean-field scenario, there is no finite critical $U$ for the
Mott transition. 

The filling-controlled Mott transition can be characterized by the charge susceptibility 
(``compressibility'')
\begin{equation}
  \kappa = \frac{\partial n}{\partial \mu} \: .
\label{eq:kappa}
\end{equation}
At $U=0$ the compressibility is proportional to the tight-binding density of states at the 
Fermi energy and is therefore finite for all fillings, except for $n=0$ and $n=2$ because 
of the van Hove singularities at the lower and upper band edge.
For any finite $U$, the compressibility must vanish in the Mott-insulating phase as for 
$n=1$ the one-particle excitation spectrum is gapped. 
Approaching the Mott insulator from the metallic side ($n \to 1$), however, the 
compressibility behaves discontinuously and even diverges. 
For $n<1$ but close to half-filling, it is given by \cite{UKO94}
\begin{equation}
  \kappa = \alpha \delta^{-1}
\label{eq:kappadiv}
\end{equation}
where $\delta = 1 - n$ is the hole concentration and $\alpha>0$ a $U$-dependent constant.
This implies that close to half-filling $\delta$, as a function of the chemical potential, 
approaches $\delta = 0$ with an infinite slope:
\begin{equation}
  \delta \propto \sqrt{\mu(0) - \mu(\delta)} \: .
\label{eq:divergence}
\end{equation}

Here, we note that the Mott physics in dimension $D=1$ stands in marked contrast to the 
DMFT (or $D=\infty$) scenario. \cite{GKKR96} 
DMFT predicts a Mott-insulating state with a vanishing $\kappa$ only for interaction strengths 
$U$ larger than a {\em finite} critical value $U_c$. 
Furthermore, for $U>U_c$ the compressibility is found to stay {\em finite}, $0<\kappa<\infty$, 
when approaching the Mott insulator from the metallic state off half-filling as $n \to 1$.

\section{Cluster approaches using the Lanczos technique}
\label{sec:cluster}

It is obviously interesting whether or not this qualitative difference in the physics 
of the Mott transition can be captured by means of a cluster extension of the DMFT.
This question has been tackled recently with the help of cellular DMFT (and with 
the periodized C-DMFT) for half-filling by Bolech et al.\ \cite{BKK03} and for the 
filling-dependent transition by Capone et al.\ \cite{CCK+04}
For studies at zero temperature the Lanczos technique \cite{LG93} is a powerful 
method to treat the effective cluster problem within C-DMFT and was also employed in 
Refs.\ \onlinecite{BKK03,CCK+04}.
The effective cluster Hamiltonian is given by
\begin{eqnarray}
  H' &=& -t \sum_{i=2}^{L_c} \sum_\sigma \left( 
  c^\dagger_{i\sigma} c_{i-1\sigma} + \mbox{H.c.}
  \right)
\nonumber \\  
  &-& \mu \sum_{i=1}^{L_c}\sum_\sigma n_{i\sigma}
   + U \sum_{i=1}^{L_c} \sum_{\sigma} n_{i\uparrow} n_{i\downarrow}
\nonumber \\
  &+& \sum_{i,\sigma} \sum_{\nu_i=2}^{n_s(i)} 
     \left(\varepsilon_{i\nu_i} 
     a^\dagger_{i\nu_i\sigma} a_{i\nu_i\sigma}
     + (V_{i\nu_i} c_{i\sigma}^\dagger a_{i\nu_i\sigma} 
     + \mbox{H.c.})
     \right)
     \: .
\nonumber \\     
\label{eq:hp}
\end{eqnarray}
Here $n_s(i)-1$ is the number of uncorrelated sites per correlated site $i$.
$n_s(i)$ may vary along the chain. 
Due to the exponential growth of the Hilbert space dimension with the system size,
calculations are basically limited to a total number of $\sum_{i=1}^{L_c} n_s(i) \sim 10$ 
sites only, if $L_c$ denotes the number of correlated sites (the cluster ``size'').
Note that the reference system has to be solved repeatedly to find a stationary point or to
achieve self-consistency, respectively, and that due to open boundary conditions a few
general symmetries can be exploited only.
For a small cluster with $L_c=4$, for example, this implies a limitation to less than 
$n_s=3$ local degrees of freedom, i.e.\ less than two bath sites per correlated site,
if $n_s(i)=n_s$ is taken to be constant as usual. 
While a true solution of the C-DMFT self-consistency equation actually requires a 
continuum of bath sites (at each correlated site), i.e.\ $n_s=\infty$, the convergence 
with respect to $n_s$ is expected \cite{CK94} to be exponentially fast. 
This makes calculations with small $n_s$ feasible. 

However, there are two conceptual drawbacks of the exact-diagonalization (Lanczos) 
approach to C-DMFT: 
(i) Clearly, the determination of the bath parameters $\varepsilon_{i\nu}$, $V_{i\nu}$ 
is of crucial importance for small $n_s$ (and small $L_c$).
One possible prescription is to fix the parameters by minimization of a suitably 
defined distance between the hybridization function of Eq.\ (\ref{eq:hp}) and the 
one given by the self-consistency equation. 
The choice of the quantity that is ``projected'' as well as the distance measure,
however, are more or less {\em ad hoc} and by no means unique. 
(ii) Within the C-DMFT the one-particle energies of and the hopping between the 
correlated sites are fixed by their values in the original Hubbard model.
This may be seen as a limited flexibility for the determination of the (in a certain
sense) optimal effective cluster model.

The variational cluster approximation (VCA) \cite{PAD03} or, more generally, the 
self-energy-functional theory \cite{Pot03a} does not suffer from these shortcomings:
(i) The bath parameters of the effective cluster model (the ``reference system'') 
are fixed in a unique way by demanding the grand potential of the system 
to be stationary with respect to those variations of the 
self-energy that are induced by varying the bath parameters.
This prescription is distinguished by the fact that it ensures thermodynamical
consistency of the results: \cite{AAPH06b,OBP07}
All approximate quantities of the theory derive from an approximate but explicitly
given thermodynamical potential. 
Opposed to C-DMFT/ED this consistency is achieved for any $L_c,n_s$ and not only in
the continuum limit $n_s \to \infty$.
(ii) There is more flexibility in the choice of the reference system: 
Within the SFT it is possible to vary {\em all} one-particle parameters of the 
reference system including those referring to the original correlated sites.
Furthermore, one is by no means forced to attach a bath to each of the correlated
sites.
A physically motivated choice is to consider bath sites at the cluster boundaries
only, for example.

\section{Variational cluster approach using $Q$ matrices}
\label{sec:sft}

The SFT is described in Refs.\ \onlinecite{PAD03,Pot03a,AAPH06b,OBP07,Pot05,Pot06a}. 
The main idea is to express the grand potential of the original model as a 
functional of the self-energy, $\Omega=\Omega[\ff \Sigma]$, such that the
exact self-energy is given as a stationary point, $\delta \Omega / \delta \ff \Sigma = 0$.
Trial self-energies are taken from a reference system with the same (Hubbard) interaction
but with a modified one-particle part. 
If the Hamiltonian of the original system, $H = H_0(\ff t) + H_1(\ff U)$, consists of a 
free part with parameters $\ff t$ and an interaction term with parameters $\ff U$, the
most general Hamiltonian of the reference system has the form 
$H'= H_0(\ff t') + H_1(\ff U)$ with arbitrary
$\ff t'$. 
Fig.\ \ref{fig:refsys} shows the original one-dimensional Hubbard model with 
nearest-neighbor hopping $t$ as well as various reference systems considered 
for our calculations.
The trial self-energy is parametrized by the set of one-particle parameters of 
the reference system: $\ff \Sigma = \ff \Sigma(\ff t')$, and variations of the trial
self-energy are considered that are due to variations of $\ff t'$, i.e.\ one has
to solve:
\begin{equation}
  \frac{\partial \Omega[\ff \Sigma(\ff t')]}{\partial \ff t'} \stackrel{!}{=} 0 \: .
\end{equation}
The decisive point is that $\Omega[\ff \Sigma(\ff t')]$ can be evaluated exactly for
reference systems that allow for a (numerically) exact computation of the single-particle
Green's function. 
In case of a finite (small) cluster or chain and a finite (small) number of bath sites, 
this can be achieved by full diagonalization or with the help of the Lanczos method. \cite{LG93}

All what is needed in a practical calculation is the one-particle Green's function of 
the reference system. 
If $L_c>1$, i.e.\ in case of the variational cluster approximation (VCA), this is the 
Green's function of a set of decoupled clusters. 
The Green's function for a single cluster,
\begin{equation}
  G'_{\alpha\beta}(\omega) = \sum_m
  Q_{\alpha m} \frac{1}{\omega - \omega'_m} Q^\dagger_{m\beta} \; ,
\end{equation}
is given in terms of poles $\omega'_m$ and corresponding weights 
$Q_{\alpha m} Q^\dagger_{m\beta}$. 
The poles and the $Q$-matrices \cite{AAPH06b} can be read off from the standard 
Lehmann representation. \cite{FW71}
Note that $\ff Q$ is a non-quadratic matrix: 
$\alpha = (i,\sigma)$ refers to a one-particle orbital of the cluster 
while $m=(r,s)$ refers to a single-particle
excitation between two eigenstates $|s\rangle$ and $|r\rangle$ of the cluster Hamiltonian
$H'$ with excitation energy $\omega'_m = E'_r - E'_s$.
We have $\ff Q \ff Q^\dagger = \ff 1 \ne \ff Q^\dagger \ff Q$.

For $T=0$ the SFT grand potential is then given by:
\begin{equation}
  \Omega[\ff \Sigma(\ff t')] = \Omega'
  + \sum_m \omega_m \Theta(-\omega_m) 
  - \sum_m \omega^\prime_m \Theta(-\omega^\prime_m) \: .
\label{eq:sftomega}
\end{equation}
Here $\Omega'$ is the grand potential of the reference system, $\Theta(\omega)$ is the 
Heaviside step function, and $\omega_m$ are the poles of the (VCA approximation for the) 
one-particle Green's function of the original system.
They can easily be obtained \cite{AAPH06b} as the eigenvalues of the matrix
\begin{equation}
  \ff M = \ff \Lambda + \ff Q^\dagger \ff V \ff Q \: ,
\label{eq:mmat}
\end{equation}
with $\Lambda_{mn} = \omega^\prime_m \delta_{mn}$ and $\ff V = \ff t - \ff t^\prime$.
Typically, $\ff V$ includes the inter-cluster hopping, shifts of one-particle energies
and, in the case of bath sites, further hybridization terms.

If a reference system with bath sites is considered, it is convenient to formally include
these bath sites also in the original system where they are, however, completely decoupled
from the correlated sites such that physical quantities remain unchanged.
This has the advantage that $\ff t$ and $\ff t'$ have the same matrix dimension, and also 
the Hamiltonians of the original and of the reference system, $H$ and $H'$, operate on the 
same Hilbert space.

The dimension of $\ff M$ is given by the number of poles of $\ff G^\prime$ with non-vanishing 
spectral weight.
If this number is not too large, the $Q$-matrix technique is a very simple means for the 
evaluation of the self-energy functional.
As there is no frequency integration involved, neither on the real axis where an additional
broadening parameter must be used, nor on the imaginary axis where a high-frequency cutoff
must be introduced and the remaining tail must be controlled, the method is also very
accurate.

For larger clusters not accessible to full diagonalization, we employ the band Lanczos method 
to compute $\ff \Lambda$ and $\ff Q$. \cite{Fre00} 
This variant of the Lanczos algorithm ensures that different elements $G'_{\alpha\beta}(\omega)$ 
have the same set of poles, i.e.\ the same $\omega'_m$ independent of $\alpha,\beta$.
The dimension of the matrix $\ff M$ is then given by the number of iteration steps
in the Lanczos procedure. 
As the results usually converge very fast, it is sufficient to consider about 100
steps only. 
This is regularly checked in our calculations.
For small clusters the results have been compared with those obtained by full 
diagonalization and found to agree within numerical accuracy.

For a given reference system one should in principle vary all one-particle parameters 
$\ff t'$ to get the optimal result.
It is much more convenient, however, to restrict oneself to a small number of physically 
motivated parameters to be optimized. 
This avoids complications arising from a search for a stationary point in a high-dimensional 
parameter space.
In most cases, as will also be demonstrated below, it is fully sufficient to consider
a few variational parameters only which are suggested by the geometry of the reference 
system in an obvious way.
The reference systems considered here as well as the corresponding variational parameters
taken into account are shown in Fig.\ \ref{fig:refsys}.

The system of interest is the $D=1$ Hubbard model. 
However, for practical purposes it is more convenient to consider a Hubbard chain consisting 
of a finite number of sites $L$ with periodic boundary conditions as our original system.
For the actual calculations we used $L = 1000 - 2000$ sites.
This is fully sufficient to ensure that all results shown below are independent of $L$.

Stationary points are determined using different numerical strategies: \cite{PTVF07}
One-dimensional parameter optimization is performed by iterative bracketing of maxima 
and minima.
For more than one variational parameter, the SFT grand potential is usually not extremal but 
has a saddle point. 
Given a certain characteristic of the saddle point, this can be found by iterated 
one-dimensional optimizations -- a strategy that has been found to be useful for
two or three parameters.
In case of higher-dimensional parameter spaces, the downhill simplex method is applied to find
local minima of $|\partial \Omega[\ff \Sigma(\ff t')] / \partial \ff t'|^2$ from which (if there is more than 
one) only those are retained for which $\Omega[\ff \Sigma(\ff t')]$ has a vanishing gradient.
For most situations \cite{Pot06a} the minimal grand potential distinguishes the 
thermodynamically stable phase if there is more than a single stationary point.
In all examples discussed below, however, this has not been an issue or turned out 
to be straightforward.

\section{Mott-insulating phase for half-filling}
\label{sec:half}

\begin{figure}[t]
\centering
\includegraphics[width=0.9\columnwidth]{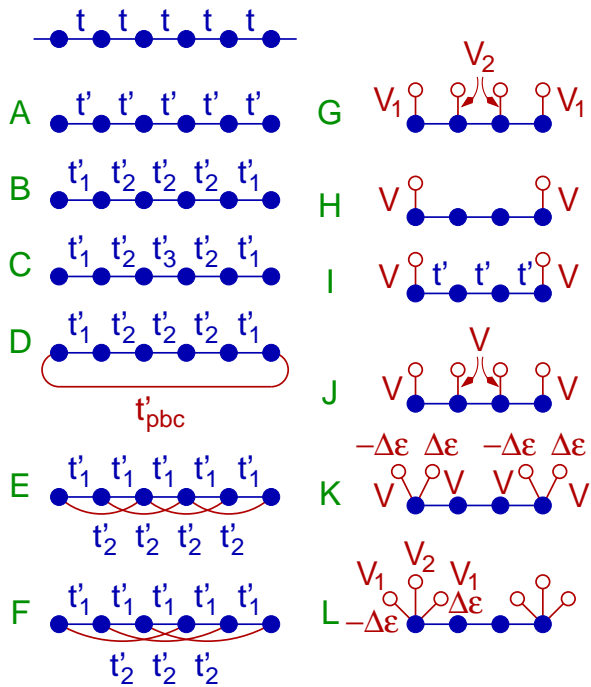}
\caption{(Color online) 
The original system ($D=1$ Hubbard model, n.n.\ hopping) and different 
reference systems considered in this study together with the corresponding variational 
parameters being optimized. See text for discussion.
}
\label{fig:refsys}
\end{figure}

One of the advantages of the variational cluster approximation (and of the SFT 
in general) consists in its flexibility to construct approximations of different
quality and complexity.
The most simple case is given by a reference system consisting of decoupled 
clusters (chains) with $L_c$ correlated sites each where only the intra-cluster 
nearest-neighbor hopping $t'$ is considered as a variational parameter, see 
Fig.\ \ref{fig:refsys}, A. 
The hopping is assumed to be the same for all pairs of nearest neighbors.
This implies that the parameter space is one-dimensional, and an {\em extremum}
of $\Omega(t') \equiv \Omega[\ff \Sigma(t')]$ defines the physical self-energy 
within this approximation.
Clusters with an even number of sites $L_c$ are preferred to avoid a Kramers-degenerate 
ground state (for half-filling) and odd-even effects when comparing results for 
different $L_c$ with each other.

Another variational parameter suggesting itself is the on-site energy. 
A homogeneous shift of all intra-cluster on-site energies acts like a separate
cluster chemical potential. 
As has been discussed in Ref.\ \onlinecite{AAPH06b}, this is actually one of the 
most important parameters as its optimization guarantees thermodynamical consistency
with respect to the particle number. 
Here, we start our discussion with the Mott-insulating phase at half-filling 
($\mu = U / 2$). 
In this case, the optimal value of the cluster on-site energies is already predicted by
particle-hole symmetry and is given by the on-site energy of the sites in the 
original model (which is set to zero).
This can also be reproduced explicitly within the VCA: 
It turns out that $\Omega[\ff \Sigma(\ff t')]$ 
is always at a maximum for vanishing on-site energies. 

We also ignore fictitious symmetry-breaking fields coupling to the local
spin or the local charge density. 
Since those coupling terms belong to the one-particle part of the Hamiltonian, 
the corresponding field strengths can in principle be considered as additional 
variational parameters. \cite{DAH+04}
A finite value of the optimized field would indicate long-range spin or charge 
order which, however, is absent in one dimension or, as concerns e.g.\ 
ferromagnetism, is disregarded here.
As discussed in Ref.\ \onlinecite{DAH+04}, the absence of antiferromagnetic order
in the $D=1$ Hubbard model at half-filling is respected by the VCA for sufficiently
large clusters.

\subsection{Variation of hopping parameters}

\begin{figure}[t]
\centering
\includegraphics[width=0.8\columnwidth]{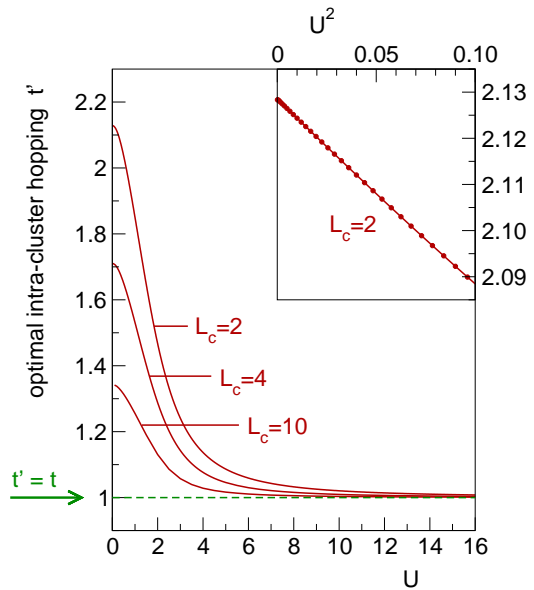}
\caption{(Color online) 
$U$-dependence of the optimal intra-cluster hopping $t'$ for different cluster 
sizes $L_c$ as indicated. 
VCA calculations for $\mu = U / 2$ (half-filling) using the reference system 
displayed in Fig.\ \ref{fig:refsys}, A. 
The physical n.n.\ hopping $t=1$ sets the energy scale.
Inset: Optimal $t'$ as function of $U^2$ for $U\to 0$.
}
\label{fig:tprime}
\end{figure}

A non-trivial result, namely $t' \ne t$, is found when optimizing the nearest-neighbor
hopping, see Fig.\ \ref{fig:tprime}. 
The physical idea behind this approximation is that switching off the {\em inter}-cluster
hopping, which generates the approximate self-energy, can partially be compensated for
by enhancing the {\em intra}-cluster hopping.
This is in fact seen in the figure: The optimal $t'$ is larger than the physical hopping. 
The trends found for different cluster sizes $L_c$ and for different $U$ corroborate
this interpretation:
The larger the cluster the smaller is the necessary compensation 
(see Fig.\ \ref{fig:tprime}).
Furthermore, it is reasonable that in case of a stronger interaction and thus more 
localized electrons, switching off the inter-cluster hopping is less significant. 
The strongest approximation of the self-energy is therefore generated by the smallest 
cluster ($L_c=2$) and in the limit $U\to 0$. 
This is indicated by a strong (more than 100\%) enhancement of $t'$ compared to $t$.

\begin{figure}[t]
\centering
\includegraphics[width=0.7\columnwidth]{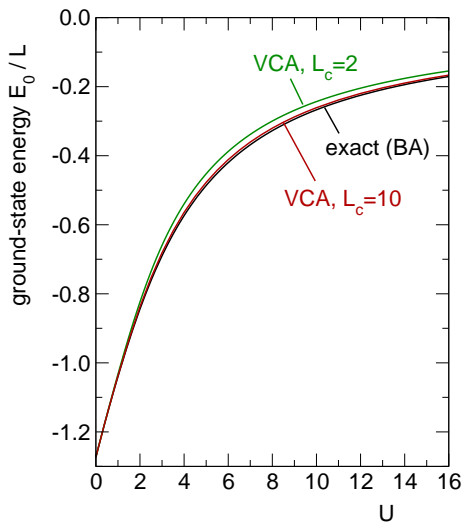}
\caption{(Color online) 
$U$-dependence of the VCA ground-state energy per site for 
cluster size $L_c=2$ and $L_c=10$ (see Fig.\ \ref{fig:refsys}, A). 
The exact (Bethe ansatz) result of Eq.\ (\ref{eq:gse}) is shown for
comparison.
}
\label{fig:gse}
\end{figure}

On the other hand, even a ``strong'' approximation for the self-energy (measured as 
a strong deviation of $t'$ from $t$) becomes irrelevant in the weak-coupling limit 
because the self-energy must vanish for $U=0$.
It is therefore not surprising that the VCA exactly recovers the $U=0$ limit.  
The approximate VCA Green's function, which can be expressed as 
$\ff G(\omega) = ({\ff G_0(\omega)}^{-1} - \ff \Sigma(\omega))^{-1}$ in terms of the
optimized self-energy and the free lattice Green's function, becomes exact for $U=0$.
The same holds for the SFT grand potential at the stationary point $\Omega$ or for the 
ground-state energy $E_0 = \Omega + \mu \langle N \rangle$. 
The latter is shown in Fig.\ \ref{fig:gse} as a function of $U$ in comparison with the 
exact (Bethe ansatz) result of Eq.\ (\ref{eq:gse}).
Both, VCA calculations for the smallest ($L_c=2$) as well as for much larger ($L_c=10$) 
cluster size, correctly reproduce the $U=0$ limit while for strong interactions there 
are deviations. 
As expected the $L_c=10$ calculation provides a much better approximation.

\begin{figure}[b]
\centering
\includegraphics[width=0.7\columnwidth]{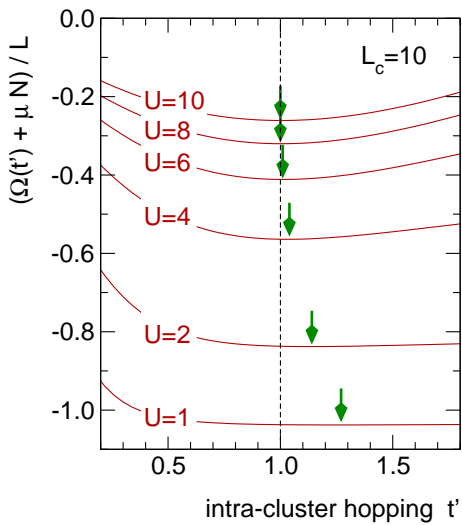}
\caption{(Color online) 
SFT grand potential $\Omega(t') \equiv \Omega[\ff \Sigma(t')]$ (constantly
shifted by $\mu \langle N \rangle$) per site and as a function of the 
intra-cluster nearest-neighbor hopping for $L_c=10$ and different $U$ 
($\mu=U/2$). 
Arrows indicate the respective optimal $t'$.
}
\label{fig:omegat}
\end{figure}

\begin{figure}[t]
\centering
\includegraphics[width=0.85\columnwidth]{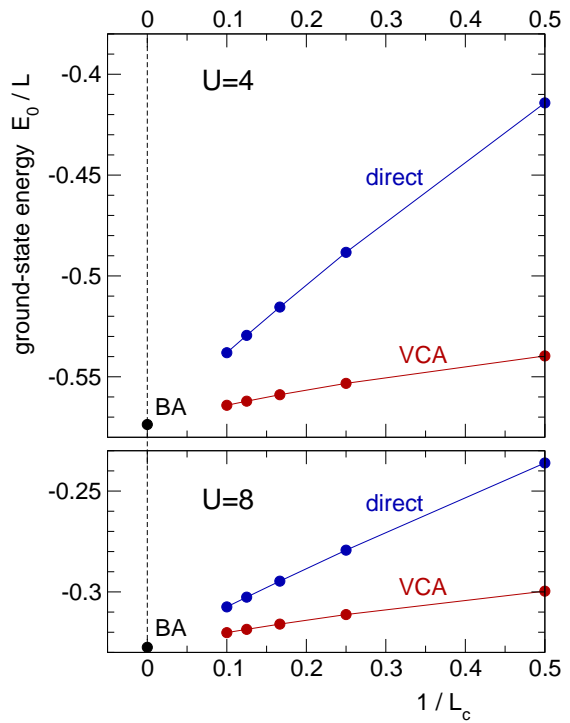}
\caption{(Color online) 
VCA ground-state energy per site for $U=4$ (top) and $U=8$ (bottom) for different
cluster sizes $L_c$ as a function of $1/L_c$ compared to the exact (BA) result and
the direct cluster approach.
}
\label{fig:u48}
\end{figure}

Fig.\ \ref{fig:omegat} demonstrates how the $U=0$ limit is approached. 
For strong interaction $U=10$ the SFT grand potential 
$\Omega(t')\equiv\Omega[\ff \Sigma(t')]$ is at a minimum for $t' \approx t$. 
Upon decreasing $U$, the optimal $t'$ more and more deviates from the physical $t$.
At the same time, however, the SFT grand potential $\Omega(t')$ becomes flatter and 
flatter, and for $U\to 0$ the optimal $t'$ is completely irrelevant as 
$\ff \Sigma(\omega) \equiv 0$ for any $t'$.

For finite $U$ the quality of the cluster approximation is 
determined by the cluster size $L_c$. 
The dependence of the VCA ground-state energy $E_0$ on $L_c$ turns out to be quite
regular. 
Plotting the results for fixed $U$ as a function of $1/L_c$ allows to recover the 
exact ground-state energy by extrapolation to $1/L_c = 0$. 
This is demonstrated in Fig.\ \ref{fig:u48}.
It is worth mentioning that the VCA represents a considerable improvement as compared
to the ``direct'' cluster approach where $E_0$ is simply approximated by the 
ground-state energy of an isolated Hubbard chain (with open boundary conditions).
Convergence to the exact result is clearly faster within the VCA. 
As can be seen by comparing the trends for $U=4$ and $U=8$ in Fig.\ \ref{fig:u48},
this advantage is more pronounced for weaker interactions which is explained by 
the fact that, opposed to the VCA, the direct cluster approach is approximate even 
for $U=0$.

Cluster-perturbation theory (CPT) \cite{GV93,SPPL00} can be understood as being 
identical with the VCA provided that the SFT expression for the grand potential 
is used and provided that isolated clusters are used as reference system and no parameter 
optimization at all is performed. 
(It should be mentioned that this implies the inability of CPT to describe antiferromagnetic
order for $D=2$ and $T=0$, for example).
As can be seen from Fig.\ \ref{fig:omegat}, there is a gain in binding energy due 
to the optimization of $t'$, i.e.\ $\Omega(t') < \Omega(t)$ for the optimal $t'$.
This means that the VCA improves on the CPT result.
One should note, however, that on the energy scale used in Fig.\ \ref{fig:u48}, for
example, this binding-energy gain would hardly be visible.

The VCA value for $E_0$, though in Fig.\ \ref{fig:u48} always higher than the exact 
result, does {\em not} represent an upper bound to the exact ground-state energy 
{\em a priori}. \cite{Pot06a}
To our knowledge there is no {\em general} proof that the self-energy functional 
is convex or ``variational'' despite several recent efforts. \cite{Kot99,CK01,NST07}
This must be seen as a disadvantage as compared to the direct cluster method which, 
via the Ritz variational principle and in the case of open boundary conditions is 
easily shown to provide strict upper bounds.
However, this disadvantage appears to be inherent to all hitherto known variational 
principles that are not derived from the Ritz principle.

\begin{figure}[t]
\centering
\includegraphics[width=0.85\columnwidth]{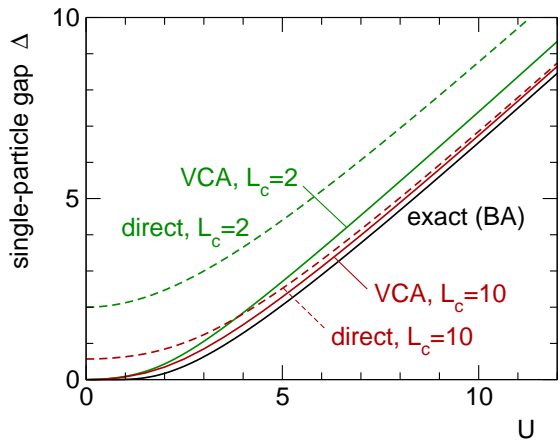}
\caption{(Color online) 
$U$-dependence of the insulating gap in the one-particle excitation spectrum 
as obtained from the VCA and from the direct cluster approach for $L_c=2$ and 
$L_c=10$ in comparison with the exact result of Eq.\ (\ref{eq:gap}).
VCA calculations using the reference system shown in Fig.\ \ref{fig:refsys}, A.
}
\label{fig:gap}
\end{figure}

The VCA derives from a dynamical variational principle based on the one-particle
self-energy as the basic variable. 
One should therefore expect that the approach is able to predict one-particle 
excitations significantly better than the direct cluster method. 
Here, for the discussion of the Mott insulator, the focus is on the insulating 
single-particle excitation gap $\Delta$, the exact $U$-dependence of which is 
given by Eq.\ (\ref{eq:gap}).
Using the $Q$-matrix approach, we get the poles of the one-particle Green's 
function with finite spectral weight by diagonalization of the matrix $\ff M$
in Eq.\ (\ref{eq:mmat}). 
The difference between the lowest pole in the electron-addition part and the 
highest pole in the electron-removal part of the spectrum defines $\Delta$.
As can be seen in Fig.\ \ref{fig:gap}, the VCA results for $L_c=2$ and $L_c=10$ 
considerably improve upon the results of the direct cluster method.
For intermediate and strong couplings, the VCA gap calculated for $L_c=10$ is
close to the exact result, and taking into account the $L_c=2$ calculation in 
addition, a finite-size scaling appears to be possible. 
In the weak-coupling limit ($U \lesssim 2$), however, an increase of the cluster
size apparently does no longer lead to a significant improvement.
Although the VCA gap approaches $\Delta=0$ for $U \to 0$, there is a clear 
overestimation as compared to the exact result with a relative error that even
diverges for $U\to 0$.

For a more detailed discussion of the critical point $U=0$ consider the inset 
in Fig.\ \ref{fig:tprime}.
One can see that for $U\to 0$, the optimal intra-cluster hopping {\em quadratically} 
approaches a finite value: $t' = t'_0 + \mbox{const.} \times U^2 + {\cal O}(U^3)$.
This implies that cluster eigenenergies and thus excitation energies as well as 
cluster eigenstates and thus spectral weights depend (for $L_c < \infty$) analytically 
on $U$ for $U \to 0$. 
Consequently, the same holds for the VCA Green's function since this can be 
expressed in terms of the cluster Green's function $\ff G'$ as 
$\ff G(\omega) = ({\ff G'(\omega)}^{-1} - (\ff t - \ff t'))^{-1}$ and since the 
matrix inversion 
involves finite blocks only due to the remaining superlattice translational symmetry
of the reference system. 
One-particle correlation functions, the ground-state energy etc.\ are therefore 
analytical in $U$ for $U\to 0$ within the VCA.
The same holds for the one-particle excitation gap while the exact gap 
is non-analytic at $U=0$ (cf.\ Eq.\ (\ref{eq:gap})).
That this non-analyticity cannot be reproduced within the VCA, should be interpreted
as a rather general failure that is inherent to any cluster concept.
Qualitative changes at a critical point resulting from the limiting process $L_c\to\infty$
are beyond a scheme based on finite clusters.

So far the discussion has been restricted to calculations using a single variational
parameter. 
More parameters can be useful for different reasons. 
First, we note that the optimal self-energy provided in a real-space cluster technique
does not reflect the full translational symmetry of the original lattice problem and that 
finite-size effects are expected to be the most pronounced at the cluster boundary.
This suggests to use reference systems with site- or bond-dependent variational parameters.
For the case of particle-hole symmetry, obvious choices are displayed in 
Fig.\ \ref{fig:refsys}, B where the intra-cluster hopping at the edges of the chain are
allowed to take a different value, and in Fig.\ \ref{fig:refsys}, C where more or all
hopping parameters are varied independently. 

\begin{figure}
\centering
\includegraphics[width=0.7\columnwidth]{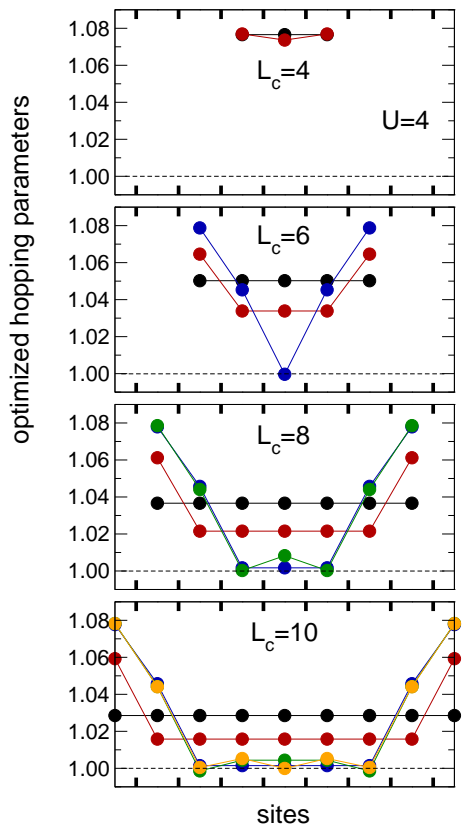}
\caption{(Color online) 
Optimized hopping parameters for reference systems shown in Fig.\ \ref{fig:refsys}, A, B and C
for $U=4$ and different cluster sizes ranging from $L_c=4$ to $L_c=10$ as indicated
(physical hopping set to $t=1$).
{\em Black}: hopping assumed to be uniform (A).
{\em Red}: two hopping parameters varied independently,
the hopping at the two cluster edges and the ``bulk'' hopping (B).
{\em Blue}: hopping at the edges, next to the edges and bulk hopping varied.
{\em Green}: four hopping parameters varied.
{\em Orange}: five hopping parameters varied (for $L_c=10$ this corresponds to C).
}
\label{fig:hopp}
\end{figure}

Fig.\ \ref{fig:hopp} shows the numerical results for $U=4$. 
We find that the optimal hopping varies between different nearest neighbors within a 
range of less than 10\%. 
At the chain edges the optimal hopping is enhanced to compensate the loss of itinerancy
due to the switched-off inter-cluster hopping within the VCA.
With increasing distance to the edges, the hopping quickly decreases.
Quite generally, the third hopping parameter is already close to the physical hopping $t$. 
Looking at the $L_c=10$ results where all (five) different hopping parameters have been
varied independently (orange circles), one can see the hopping to slightly oscillate around 
the bulk value reminiscent of surface Friedel oscillations. 

The optimal SFT grand potential is found to be lower for the inhomogeneous cases as compared
to the homogeneous (black) one.
Generally, the more variational parameters are taken into account the higher is the decrease
of the SFT grand potential at optimal parameters.
However, the binding-energy gain due to inhomogeneous hopping parameters is much smaller 
compared to the gain obtained with a larger cluster. 
Likewise, there is merely a marginal improvement as concerns the single-particle gap.

Considering an additional hopping parameter $t_{\rm pbc}$ linking the two chain edges as 
a variational parameter (Fig.\ \ref{fig:refsys}, D), always gives a minimal SFT grand potential
at $t_{\rm pbc}=0$. 
This implies that open boundary conditions are preferred as compared to periodic boundary
conditions (which would be given by a stationary point at $t_{\rm pbc} = 1$). 
The issue has already been discussed in Ref.\ \onlinecite{PAD03}.

With the reference system Fig.\ \ref{fig:refsys}, E we can check whether or not a magnetic 
frustration develops in the reference system. 
A hopping $t'_2$ between next-nearest neighbors leads in the Heisenberg limit $U \to \infty$
to an antiferromagnetic next-nearest-neighbor exchange $J_2$ and thus to a frustration of 
antiferromagnetic (short-range) order. 
This would partially compensate the residual mean-field character of the VCA with respect to
magnetic properties.
At the same time, however, particle-hole symmetry would be violated.
It turns out, however, that the SFT grand potential has a saddle point with $t_2'=0$.
(It is at a minimum w.r.t.\ $t_1'$ and at a maximum w.r.t.\ $t_2'$).

A third-nearest-neighbor hopping would not lead to frustration and would also respect 
particle-hole symmetry. 
Optimization of an $L_c=6$-site cluster at $U=4$ as indicated in Fig.\ \ref{fig:refsys}, F
yields an optimal nearest-neighbor hopping $t_1' \approx 1.04$ and third-nearest-neighbor 
hopping $t_2' \approx -0.02$.
This shows that hopping parameters that are not present in the original system can get a
finite value when treated as variational parameters in the reference system. 
The corresponding decrease of the SFT grand potential is marginal, however. 
Consequently, we disregard such variational parameters in the following. 

\subsection{Bath degrees of freedom}

A different possibility to increase the number of variational parameters is to introduce
additional uncorrelated (``bath'') sites.
As there is no Hubbard interaction on the bath sites, the interaction part of the Hamiltonian is
left unchanged, as it is necessary for an allowed reference system within the SFT. \cite{Pot03a}
Note that the trial self-energy $\Sigma_{ij}(\omega)$ is still labelled by the correlated sites only. 
We consider reference systems where all or some of the original correlated sites are coupled
to bath sites via a hopping (``hybridization'') of strength $V$.
For each correlated site $i$ the different hybridization parameters $V_{i\nu_i}$ as well as the 
one-particle energies of the bath sites $\varepsilon_{i\nu_i}$ for $\nu_i = 2,...,n_s(i)$ can be treated 
as variational parameters.
Here $n_s(i)-1$ is the number of bath sites for a given correlated site $i$.
The inclusion of bath sites improves the description of temporal instead of spatial degrees
of freedom.
For $L_c=1$ one recovers the DMFT, for $L_c>1$ the cellular DMFT in the limit $n_s(i)=n_s\to\infty$.
\cite{PAD03}
Calculations using the Lanczos method are feasible, however, for small $n_s$ only.

Particle-hole symmetry considerably reduces the number of variational parameters that have
to be varied independently. 
For a single bath site ($n_s(i)=2$), the bath on-site energy is pinned to the chemical potential, 
$\varepsilon = \mu = U/2$, and only the hybridization $V$ is free.
For $n_s(i)=3$ we have $\varepsilon = \mu \pm \Delta \varepsilon$ with a variational parameter
$\Delta \varepsilon$. 
Both bath sites couple with the same $V$ to the correlated site. 

\begin{figure}
\centering
\includegraphics[width=0.9\columnwidth]{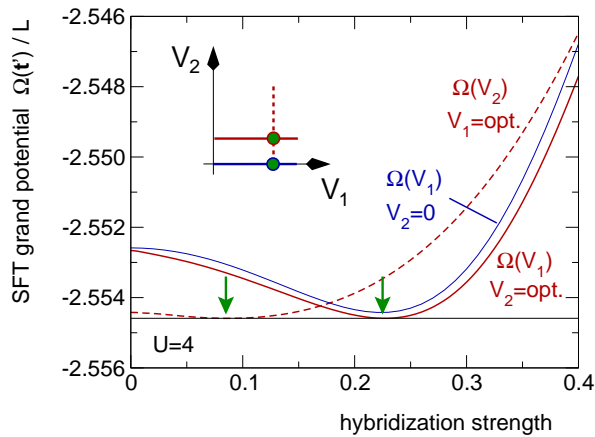}
\caption{(Color online) 
SFT grand potential per site $\Omega(\ff t') / L \equiv \Omega[\ff \Sigma(\ff t')] / L$ 
as a function of the hybridization strength for $U=4$.
Red lines: reference system Fig.\ \ref{fig:refsys}, G; solid: $\Omega(V_1)$ at $V_2=\mbox{opt.}$,
i.e.\ varying $V_1$ of the edge bath sites while keeping the hybridization of the central bath 
sites $V_2$ at its optimal value.
Dashed: $\Omega(V_2)$ at $V_1=\mbox{opt.}$.
Arrows indicate the respective minima.
Blue line: reference system Fig.\ \ref{fig:refsys}, H, i.e.\ $\Omega(V_1)$ at $V_2=0$.
}
\label{fig:fullbath}
\end{figure}

For a cluster approximation with $L_c > 1$, baths should be different for inequivalent 
correlated sites. 
It has to be expected, for example, that bath sites at the cluster boundary are more 
efficient to compensate for the disregarded inter-cluster hopping processes than bath sites
coupled to the cluster center.
This can be studied using the reference system G in Fig.\ \ref{fig:refsys} which includes
a coupling to a bath site at the edges ($V_1$) and at the central sites ($V_2$) of a cluster
with $L_c=4$ correlated sites.
Fig.\ \ref{fig:fullbath} shows the results of the according VCA calculation. 
Both the inner and the outer bath sites couple to the system with a finite hybridization
and thereby lead to a decrease of the optimal SFT grand potential 
as compared to vanishing hybridization.
For the outer ones, however, this binding-energy gain is about one order of magnitude higher.
Also the optimal hybridization is much larger for the outer bath sites.

This suggests to neglect the coupling of bath sites to the correlated sites at the center of 
the chain completely, i.e.\ to switch off $V_2$ and to consider reference system H in 
Fig.\ \ref{fig:refsys}. 
As can be seen in Fig.\ \ref{fig:fullbath}, this represents an excellent approximation.
Comparing the results for the reference systems G and H with each other by looking at the
trend of $\Omega(V_1)=\Omega[\ff \Sigma(V_1)]$ for optimal $V_2$ and for $V_2=0$, 
respectively, we find the optimal SFT grand potential to be only slightly higher 
and the optimal hybridization $V_1$ almost unchanged.

\begin{figure}
\centering
\includegraphics[width=0.8\columnwidth]{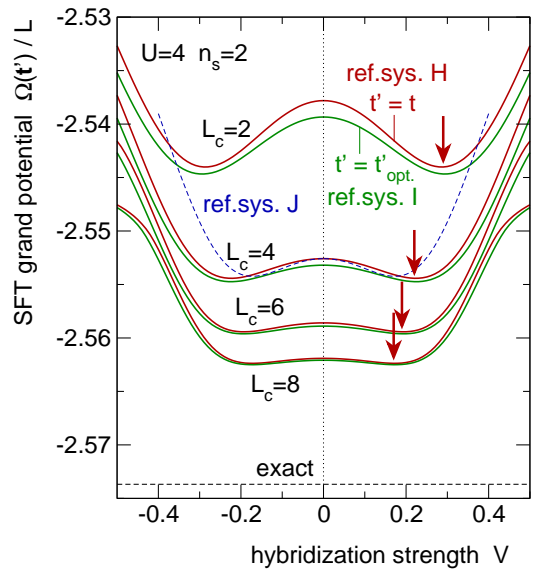}
\caption{(Color online) 
SFT grand potential per site as a function of the hybridization strength for $U=4$.
Red lines: reference system Fig.\ \ref{fig:refsys}, H for different cluster sizes ranging 
from $L_c=2$ to $L_c=8$.
Green lines: reference system Fig.\ \ref{fig:refsys}, I, i.e.\ with additional optimization
of the nearest-neighbor hopping.
Arrows indicate minima. 
Dashed black line: BA result for $E_0 - \mu N$.
}
\label{fig:bathedge}
\end{figure}

The idea of attaching bath sites at the cluster edges only is pursued with the calculations
shown in Fig.\ \ref{fig:bathedge}.
We employ reference system H.
For any cluster size from $L_c=2$ to $L_c=8$, it is found that edge bath sites couple to the 
system and decrease the SFT grand potential.
For $L_c=4$ and $U=4$ (see figure) this decrease amounts to $\Delta \Omega /L \approx 0.002$.
For stronger interactions the cluster approximation generally tends to improve.
Consequently, the optimal hybridization becomes smaller.
The optimal grand potential at $U=8$, for example, decreases by $\Delta \Omega / L \approx 0.0001$
due to the bath sites which is one order of magnitude less than for $U=4$. 
It must be emphasized that thanks to the $Q$-matrix technique, \cite{AAPH06b} which completely avoids 
frequency summations or integrations, there are no numerical problems to accurately compute energy 
differences of this or even lower order of magnitude.
From the numerical point of view, this is an important step forward as compared to earlier evaluations
of the SFT grand potential using integrations over frequencies $\omega + i \eta$ with real $\omega$ 
and subsequent extrapolation $\eta\to 0$ (cf.\ Refs.\ \onlinecite{PAD03,AAPH06b} for a detailed
comparison).

\begin{figure}[b]
\centering
\includegraphics[width=0.65\columnwidth]{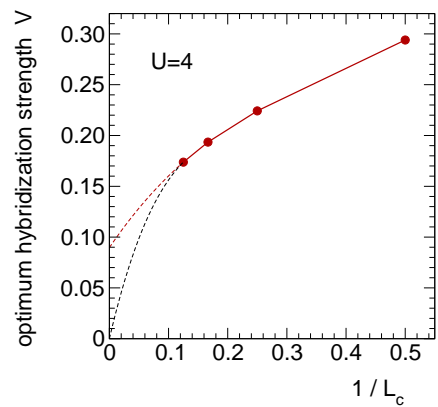}
\caption{(Color online) 
Optimal values of the hybridization in Fig.\ \ref{fig:bathedge} as a function of the inverse
cluster size.
Dashed lines show two possible extrapolations to the $L_c=\infty$ limit.
}
\label{fig:vopt}
\end{figure}

Attaching bath sites and thereby allowing the electrons to hop into the bath and back to the
original site and thereby to gain kinetic energy, turns out to be more effective than the gain
in kinetic energy that is obtained by optimizing (and increasing) the intra-cluster hopping $t'$. 
This is demonstrated in the figure by calculations using the reference system I where the hybridization
to edge bath sites and the intra-cluster hopping are optimized simultaneously.
Comparing the results for reference systems H and I shows that for any $L_c$ the binding energy gain 
due to the coupling of bath sites is considerably larger.

Fig.\ \ref{fig:bathedge} also includes the result of a calculation using reference system J (with $L_c=4$) 
where there is one bath site for {\em any} correlated site (not only at the edges) but still only a single 
hybridization strength that is optimized by assuming this to be the same for all bath sites.
It is interesting to note that this reference system turns out to be inferior as compared to I and also 
to H (the SFT grand potential at the minimum is higher) although there are two more bath sites.
This once more demonstrates the ineffectiveness of bath sites coupled to the center of the Hubbard chain. 

With increasing $L_c$ the optimal SFT grand potential (using reference system H, for example) 
nicely converges to the exact value which is shown in Fig.\ \ref{fig:bathedge} for comparison.
It is important to note that the inclusion of bath sites hardly speeds up this convergence.
For any given cluster size $L_c$, the additional inclusion of two more bath sites gives a binding-energy
gain considerably smaller than the gain obtained by a cluster with two more correlated sites. 
This also holds true if more bath sites are taken into account. 
The decisive lowering of the SFT grand potential is always due to a larger cluster size. 
Concluding, bath sites are quite ineffective as far as the grand potential or the ground-state 
energy is concerned.

It is an interesting question whether or not bath sites at the edges of the cluster finally 
decouple from the correlated sites for $L_c\to \infty$, i.e.\ whether or not the respective 
optimal $V$ vanishes in this limit.
For bath sites coupled to the center of the cluster, a decoupling $V\to 0$ is quite plausible 
physically and is actually foreshadowed by the results shown in Fig. \ref{fig:fullbath} for 
reference system G.
For edge bath sites, the optimal $V$ is shown in Fig.\ \ref{fig:vopt} as a function of the inverse
cluster size.
As a simple cubic spline extrapolation shows, the results are compatible with a finite $V$ for
$L_c\to \infty$ (red dashed line) but also $V=0$ cannot be excluded.
From the results shown in Fig.\ \ref{fig:bathedge} it appears obvious that a minimum of 
$\Omega[\ff \Sigma(\ff t')]/L$ is found for any finite $L_c$.
A finite position of the minimum for $L_c\to \infty$ would imply an interesting behavior of the 
SFT grand potential as function of $V$ as this must become completely flat (at least in a 
finite $V$ range around $V=0$).

While bath sites are of minor importance as concerns static quantities such as the ground-state
energy, they are decisive for dynamical quantities and for the single-particle excitation gap 
$\Delta$ in particular.
This shall be demonstrated in the following.
As argued above, a cluster method is likely to fail close to the critical point $U=0$.
Predicting the gap away from the critical point, however, can serve as any strong test for
a cluster technique.

Tab.\ \ref{tab:gap} shows results for the $\Delta$ at intermediate coupling $U=4$ as obtained
within different approaches. 
By evaluation of Eq.\ (\ref{eq:gap}) we find the exact value $\Delta = 1.287$.
As it is well known, a metallic state with $\Delta=0$ is predicted by static mean-field theory 
and, for $U=4$, by dynamical single-site approximations (dynamical impurity approximation \cite{Pot03a} 
and DMFT).
On the other hand, even the most simple cluster approach, i.e.\ the VCA for $L_c=2$ and 
$n_s=1$ (without bath sites), predicts a finite gap but strongly overestimates its size. 
Improvement is possible using larger clusters, but even for $L_c=10$ the gap is overestimated by 
about 18\% (see also Fig.\ \ref{fig:gap}).
This remains essentially unchanged even with a full optimization of 5 different hopping parameters
(Tab.\ \ref{tab:gap}, VCA for reference system C).

\begin{table}[t]
\begin{tabular}{l|l}
\hline
\hline
   & excitation gap $\Delta$ \\
\hline
exact, \cite{Ovc69} Eq.\ (\ref{eq:gap})    & \mbox{}\hspace{2mm} 1.287 \\ 
DMFT                                       & \mbox{}\hspace{2mm} 0.0 \\
\hline	
VCA, A, $L_c = 2$                          & \mbox{}\hspace{2mm} 1.846 \\  
VCA, A, $L_c =10$                          & \mbox{}\hspace{2mm} 1.516 \\  
VCA, C, $L_c =10$                          & \mbox{}\hspace{2mm} 1.518 \\  
\hline
VCA, H, $L_c = 2$, $n_s=2$                 & \mbox{}\hspace{2mm} 0.238 \\  
VCA, H, $L_c = 4$, $n_s=2$ (edge)          & \mbox{}\hspace{2mm} 0.079 \\  
VCA, L, $L_c = 2$, $n_s=4$                 & \mbox{}\hspace{2mm} 0.009 \\
\hline
VCA, K, $L_c = 2$, $n_s=3$                 & \mbox{}\hspace{2mm} 1.181 \\ 
VCA, K, $L_c = 4$, $n_s=3$ (edge)          & \mbox{}\hspace{2mm} 1.213 \\ 
\hline
C-DMFT \cite{BKK03}                        & \mbox{}\hspace{2mm} 1.14 \\  
\hline
\hline
\end{tabular}
\caption{
Single-particle excitation gap at $U=4$ as obtained within the VCA using different reference
systems as indicated (see Fig.\ \ref{fig:refsys}). We set $t=1$. C-DMFT result for $L_c=2$ and
$n_s=3$ (Ref.\ \onlinecite{BKK03}).
}
\label{tab:gap}
\end{table}

Introducing bath sites completely changes the situation. 
Using reference system H, i.e.\ $L_c=2$ correlated sites with one bath site attached to each 
($n_s=2$), yields a gap which is drastically too small. 
The result becomes even worse, namely the gap is almost closed, when increasing the cluster size
to $L_c=4$ but still keeping one bath site attached to each of the two edge correlated sites.
We infer that while the $n_s=2$ cluster nicely improves the ground-state energy, it apparently
fails to describe the excitation gap. 

Using one more bath site (reference system K, $n_s=3$) and $L_c=2$ yields a further but negligibly 
small decrease of the ground-state energy but a gap that comes very close to the exact one 
(which is underestimated by about 8\%).
Now, an increase of the cluster size to $L_c=4$ yields further improvement, namely a gap that 
underestimates the exact one by 6\% only.
Adding another bath site at each of the cluster edges, i.e.\ $n_s=4$ (reference system L), 
yields an almost vanishing gap again.
We conclude that there is a sizable odd/even effect with respect to the number of bath sites, 
and that a reliable prediction of the gap
requires an even number (i.e.\ $n_s$ odd). 
Given this, the inclusion of bath sites is of crucial importance for an accurate determination 
of single-particle excitations and the insulating gap.
 
We would like to stress that due to the SFT variational principle and due the Q-matrix technique 
there is no adjustable parameter in the calculation of the gap, once the reference system is 
specified.
It has been verified that the results are converged with respect to the size of the original 
$D=1$ Hubbard model ($L \sim 10^3$) and with respect to the number of Lanczos steps 
($S_L \sim 100$).
Within the cellular DMFT (using Lanczos as a cluster solver) on the other hand, the gap value 
somewhat depends on the projection criterion employed to fix the bath parameters.
In Ref.\ \onlinecite{BKK03} the definition of the C-DMFT gap is furthermore adjusted to recover the 
exact value in the strong-coupling limit. 
Thereby, a gap of $\Delta = 1.14$ is obtained for an $L_c=2$, $n_s=3$ cluster which is close to 
our result (see Tab.\ \ref{tab:gap}).

\begin{figure}
\centering
\includegraphics[width=0.95\columnwidth]{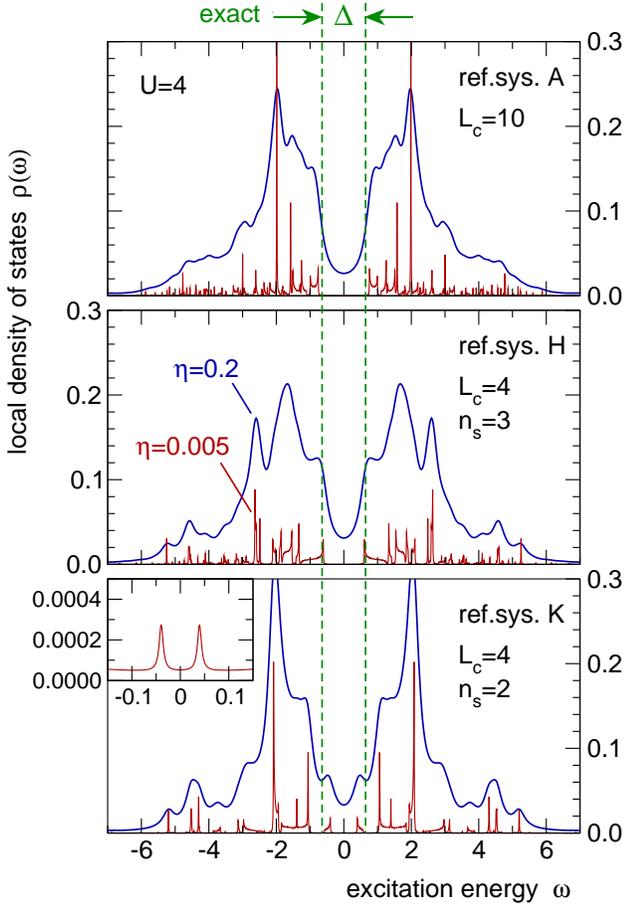}
\caption{(Color online) 
Local density of states at a central site as obtained for different reference systems 
from the VCA Green's function for $U=4$. 
Calculations are performed for two different Lorentzian broadening parameters $\eta$.
Results for $\eta=0.005$ have been scaled by a common constant factor $1/20$. 
The exact single-particle gap is indicated by dashed lines. 
}
\label{fig:dos}
\end{figure}

\begin{figure}[t]
\centering
\includegraphics[width=0.95\columnwidth]{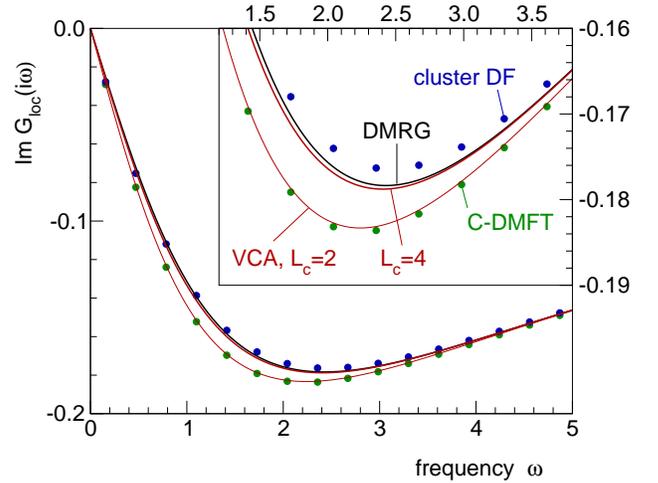}
\caption{(Color online) 
Local Green's function as a function of Matsubara frequency for $U=6$ using the 
$L_c=2$, $n_s=3$ and the $L_c=4$, $n_s=3$ reference systems (red solid lines) compared 
to dynamical DMRG data (black solid line), $L_c=2$ C-DMFT at finite (low) temperature 
($\beta=20$) and $L_c=2$ cluster dual-fermion theory (cluster DF, $\beta=20$) taken 
from Ref.\ \onlinecite{HBR+07}.
}
\label{fig:glocal}
\end{figure}

Using the reference system K with $L_c=4$ and $n_s=3$, we have computed the gap also for 
different $U$. 
The deviation to the exact result is found to decrease further for stronger interactions 
and amounts to $1.7\%
$ for $U=6$ and $0.3\%
$ for $U=8$.
The gap size is underestimated.
For smaller $U$, deviations are larger (underestimation of $10\%
$ and $7\%
$ for $U=3$ and 
$U=2$, respectively).
Compared with the C-DMFT results, \cite{BKK03} however, this is still a considerable improvement.
As a cluster approach the VCA cannot reproduce the exponentially small gap in the critical 
regime and finally {\em over}estimates the gap size with a relative error of about $340\%
$ at $U=1$ being a typical example (here the exact gap is $\Delta=0.005$).

To analyze the origin of the apparent odd/even effect, we discuss the interacting local 
density of states (LDOS) which is shown in Fig.\ \ref{fig:dos} for $U=4$.
The LDOS is calculated from the imaginary part of the local VCA Green's function which,
opposed to DMFT or C-DMFT, differs from the cluster Green's function. 
Since a real-space cluster approach necessarily breaks translational symmetry, the LDOS is
computed for a central cluster site where convergence for $L_c \to \infty$ is expected to
be the fastest.
The LDOS turns out to be non-vanishing on a large number of small but finite frequency 
intervals.
This structure is an artifact that is due to the finite Hilbert space corresponding to 
the reference system and due to the corresponding discrete pole structure of the VCA 
self-energy.
A smooth LDOS can therefore only be obtained with some additional broadening.
For the figure we have used a Lorentzian broadening with a comparatively large broadening 
parameter $\eta=0.2$ (blue lines) while the gap size (which is calculated for $\eta=0$) 
can be read off quite accurately from the spectra with small $\eta=0.005$ (red lines).

Comparing the results for the different reference systems with each other, we find the same
overall structure of the LDOS in all cases but also significant finite-size effects. 
The latter prevent a reliable prediction of the detailed shape of the LDOS.
While we expect the $L_c=10$ cluster (without bath sites, upper panel) to give the best 
estimate for the LDOS in general, the discussion above has shown that the $L_c=4$, $n_s=3$
cluster yields a much more reliable prediction of the insulating gap (middle).
Looking at the result for $\eta=0.005$, a deviation from the exact result is in fact hardly 
visible on the scale of the figure for this cluster.
Contrary, the LDOS computed for the $L_c=4$, $n_s=2$ cluster (lower panel) clearly shows 
finite spectral weight for frequencies much smaller than the exact gap (see also the inset).
This weight is rather small but significant as can be seen by varying the chemical potential:
A slight decrease from its particle-hole symmetric value $\mu=U/2$ of about 
$\Delta \mu = 0.04$ is sufficient to produce a metallic state.

To understand the failure of the $L_c=4$, $n_s=2$ approximation, one can ({\em ad hoc})
decrease the intra-cluster nearest-neighbor hopping from $t'=t$ to $t'=0$ in the reference 
system. 
This results in an LDOS with a three-peak structure consisting of the two Hubbard bands 
and a quasi-particle peak at $\omega=0$ as it is typical for the metallic solution of 
the half-filled Hubbard model within a dynamical impurity 
(mean-field) approximation given by $L_c=1$. 
The spectral weight in Fig.\ \ref{fig:dos} (lower panel) that is responsible for the too
small gap in the case of $n_s=2$ (odd number of bath sites), thereby continuously evolves 
into the quasi-particle peak of the metallic mean-field LDOS.
Now, within the dynamical impurity approximation, such a metallic LDOS at half-filling is 
always produced by an {\em odd} number of bath sites where one bath site has a one-particle 
energy exactly at the Fermi edge $\mu$ while the energies of the remaining are located 
symmetrically to $\mu$.
Due to a decreasing and eventually vanishing hybridization strength, this Fermi-edge bath 
site decouples from the rest of the reference system upon approaching the Mott insulator 
with increasing $U$.
The Mott insulator itself is therefore characterized by an {\em even} number of bath sites.
We therefore interpret the occurrence (absence) of low-frequency spectral weight within 
the exact gap as being a reminiscence of the low-frequency character of the corresponding 
metallic (insulating) mean-field solution. 
Hence, a reasonable description of the Mott insulator and a reliable prediction of the 
insulating gap requires an even number of bath sites per correlated site also in the case 
of cluster ($L_c > 1$) approximations.

Finally, we compare our results for the local Green's function on the imaginary frequency
axis with calculations by Hafermann et al.\ (see Ref.\ \onlinecite{HBR+07}) for $U=6$.
Fig.\ \ref{fig:glocal} shows our VCA results for the $L_c=2$ cluster with $n_s=3$ local 
degrees of freedom per correlated site in comparison with essentially exact dynamical 
DMRG data.
As compared with $L_c=2$ cellular DMFT using weak-coupling continuous-time QMC at finite 
temperature ($1/T = \beta = 20$), there is a marginal improvement only. 
We would like to stress, however, that the VCA calculations are computationally much cheaper.
The cluster dual-fermion approach for $L_c=2$ (and $\beta=20$) gives a considerably better 
result and is already quite close to the DMRG data.
Further improvement is possible for larger clusters.
VCA calculations for the $L_c=4$, $n_s=3$ reference system, i.e.\ with two more correlated 
sites, are hardly distinguishable from the DMRG.

\section{Filling-dependent Mott transition}
\label{sec:fill}

To complete the benchmarking of different cluster approximations, we study the metallic
phase off half-filling and the filling-dependent Mott transition. 
The metallic phase is characterized by a finite compressibility $\kappa$ (see 
Eq.\ (\ref{eq:kappa})).
Changing the electron density (filling) $n$ by changing the chemical potential $\mu$ 
at fixed $U$, 
the Mott insulator at $n=1$ is approached with a diverging $\kappa \to \infty$ for 
$n \to 1$ (see Eqs.\ (\ref{eq:kappadiv}) and (\ref{eq:divergence})).

An approximation that is based on a cluster with a finite (and small) number of degrees
of freedom, necessarily implies a strongly limited frequency resolution and thus a rough 
description of low-frequency physical properties.
We therefore expect that it is more difficult within the VCA to describe a metallic
state with gapless single-particle excitations as compared to the description of a 
gapped Mott-insulating state.
On the other hand, this argument neglects the fact that the limitation of the frequency
resolution can at least partly be compensated for by an adaption of the cluster 
one-particle parameters.
This is well known from the DMFT-ED approach \cite{CK94} as well as from dynamical 
impurity approximations \cite{Pot03a} which show that the low-frequency quasi-particle 
(Kondo) resonance can be accessed with a few sites only.
At least technically, however, VCA calculations are more difficult for the 
metallic state. 
This is simply due to the fact that, for a given cluster size, the absence of particle-hole 
symmetry implies an increased number of independent variational parameters to be optimized.

\subsection{Variation of one-particle energies}

It suggests itself, for example, to include an overall shift of the one-particle
energies of the cluster sites into the set of variational parameters.
Note that for the particle-hole symmetric case this shift is irrelevant (the VCA 
grand potential is stationary at a vanishing shift).
At a finite hole concentration, however, the grand potential is at a maximum for a
finite shift (which is different from the shift of the chemical potential). 
As has been shown in Ref.\ \onlinecite{AAPH06b}, this ensures the thermodynamical 
consistency of the approach with respect to the particle number.
Especially for the filling-dependent Mott transition, it is important that the filling 
calculated from the (approximate) interacting density of states,
\begin{equation}
  n = \frac{1}{L} \sum_{i\sigma} \int_{-\infty}^0 d\omega \: \rho_{i\sigma}(\omega) \: ,
\label{eq:ndos}
\end{equation}
where $\rho_{i\sigma}(\omega) = (-1/\pi) \mbox{Im} \, G_{ii\sigma}(\omega+i0^+)$,
gives the same result as the filling calculated from the (approximate) grand potential,
\begin{equation}
  n = - \frac{1}{L} \frac{\partial \Omega}{\partial \mu} \: ,
\label{eq:nomega}
\end{equation}
i.e.\ that calculations on the level of one-particle excitations are consistent with 
those on the (zero-particle) static thermodynamical level.

The one-particle excitation spectrum is most accurately determined by using the 
dynamical density-matrix renormalization-group (DMRG) technique \cite{BGJ04} or the 
quantum Monte-Carlo approach. \cite{PMvdL+94,ZAHS98}
The essential features of the spectrum, including the more intense spinon and 
holon bands, however, are already accessible using finite Hubbard chains of 
rather small size (e.g.\ $L_c=10$).
This has already been demonstrated by calculations using cluster-perturbation theory (CPT), 
\cite{SPPL00} i.e.\ VCA without any parameter optimization at all.
Even for $L_c=2$ and a number of $n_s - 1 = 2$ additional bath sites per correlated 
site, the overall spectrum is in very good agreement with the more accurate 
DMRG results as has been demonstrated by Capone et al.\cite{CCK+04}

\begin{figure}
\centering
\includegraphics[width=0.7\columnwidth]{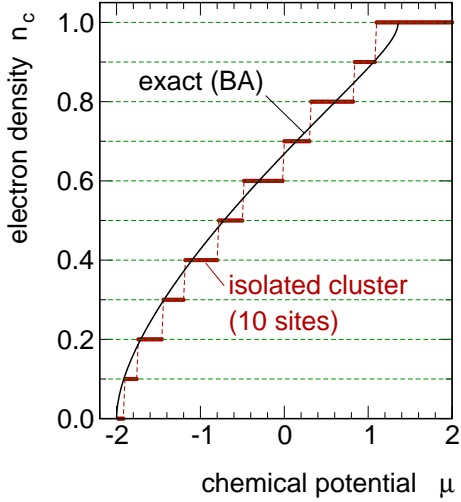}
\caption{(Color online) 
Electron density $n_c = \langle N \rangle / L_c = N_c / L_c$ as a function of the 
chemical potential for an $L_c=10$ Hubbard chain (open boundary conditions) compared 
to the exact result for $L_c \to \infty$ from Ref.\ \onlinecite{UKO94}.
}
\label{fig:edndep}
\end{figure}

\begin{figure}
\centering
\includegraphics[width=0.9\columnwidth]{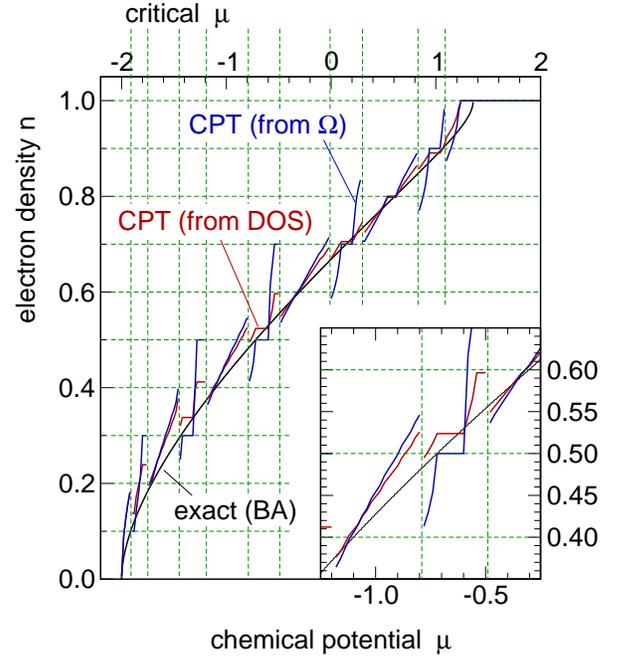}
\caption{(Color online) 
Electron density $n$ within cluster-perturbation theory (CPT), calculated by integrating 
the CPT density of states up to zero excitation energy (red line, Eq.\ (\ref{eq:ndos})) 
and calculated by differentiation of the CPT grand potential with respect to $\mu$ (blue 
line, Eq.\ (\ref{eq:nomega})), as function of $\mu$ compared to the exact result (BA).
Cluster size: $L_c=10$.
Horizontal dashed lines: cluster fillings.
Vertical dashed lines: critical chemical potentials at which the cluster ground state changes.
}
\label{fig:cptndep}
\end{figure}

\begin{figure}[b]
\centering
\includegraphics[width=0.75\columnwidth]{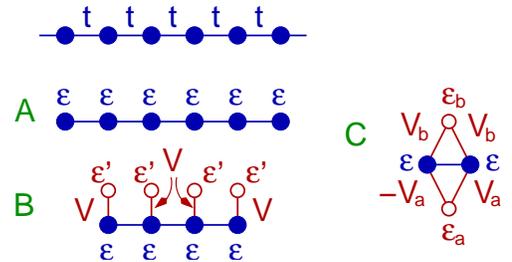}
\caption{(Color online) 
The original system and different reference systems with corresponding 
variational parameters. See text for discussion.
}
\label{fig:refsys2}
\end{figure}

Here we would like to focus on a different point which is relevant for any cluster
approach.
This is illustrated in Fig.\ \ref{fig:edndep} for an isolated cluster with $L_c=10$
sites.
Due to the U(1) symmetry of the grand-canonical cluster Hamiltonian $H'$, 
the cluster ground state has a definite total particle number $N_c$ which, depending
on the chemical potential $\mu$, can vary from $N_c=0$ to $N_c=2L_c$.
Hence, the density $n_c=N_c/L_c$ acquires discrete values only when varying $\mu$ 
and discontinuously jumps at certain critical chemical potentials $\mu_{c,i}$.
As is demonstrated by the figure, this is a strong finite-size effect that cannot be
tolerated if one is interested in the filling-dependent Mott transition of the 
system in the thermodynamical limit.

Therefore, it is an obvious question whether it is possible to predict a continuous
and reliable trend of the filling as a function of $\mu$ for $L_c \to \infty$ but 
using an approximation based on finite (small) clusters only.
This is also related to the question whether one can access systems with an arbitrarily 
low hole concentration $\delta = 1 - n$ as it is necessary, for example, to recover
the compressibility divergence for $\delta \to 0$.

By considering an infinite system of disconnected clusters and by re-introducing the 
inter-cluster hopping in the lowest non-trivial order, cluster-perturbation theory
directly works in the thermodynamical limit.
Therefore, CPT could by able to predict a continuous $\mu$ dependence, in
principle. 
As can be seen from Fig.\ \ref{fig:cptndep}, CPT in fact gives a metallic state
with a compressibility $\kappa$ that is finite everywhere except for the critical 
chemical potentials of the isolated cluster at $\mu = \mu_{c,i}$.
This is a substantial improvement as compared to the direct diagonalization 
where $\kappa \equiv 0$ (except for $\mu = \mu_{c,i}$).
However, the obvious disadvantage is that the CPT filling exhibits finite jumps 
at $\mu = \mu_{c,i}$.
This is easily understood by looking at Eq.\ (\ref{eq:ndos}) since for each 
$\mu=\mu_{c,i}$ there is a discontinuous change of the ground state of the 
(grand-canonical) cluster Hamiltonian which implies a discontinuous change of 
the cluster Green's function $\ff G'(\omega)$ and thus of the lattice (CPT) 
Green's function $\ff G(\omega) = ({\ff G'(\omega)}^{-1} - (\ff t - \ff t'))^{-1}$.

Essentially the same applies to the case where $n$ is calculated as the $\mu$ 
derivative of the grand potential, see Eqs.\ (\ref{eq:sftomega}) and (\ref{eq:nomega}).
Furthermore, the comparison of the results obtained from Eqs.\ (\ref{eq:ndos}) and 
(\ref{eq:nomega}) illustrates the thermodynamical inconsistency of the CPT
(see Fig.\ \ref{fig:cptndep}).
Compared with the exact Bethe ansatz result, it turns out to be much better 
to calculate the filling from the CPT density of states.

We conclude that plain CPT cannot describe the filling-dependent Mott transition
and exhibits severe problems in describing the trend of $n$ as a function of the 
chemical potential.
Restricting the approach to the discrete set of cluster densities, however, gives
rather satisfactory results. 
Fig.\ \ref{fig:cptndep} shows that the chemical potential, where the CPT filling 
equals one of the accessible cluster fillings, is close to the exact $\mu$ 
corresponding to that filling and that both ways to compute $n$ (Eqs.\ (\ref{eq:ndos}) 
and (\ref{eq:nomega})) almost yield the same result.
This is nicely demonstrated in the inset for $n=0.4$ and $n=0.6$.
However, there are still problems, even for an accessible cluster filling as, 
for example, in the case of $n=0.5$ where no unique chemical potential can be 
read off and $\kappa \equiv 0$ in a finite $\mu$ range (see inset again).

\begin{figure}
\centering
\includegraphics[width=0.9\columnwidth]{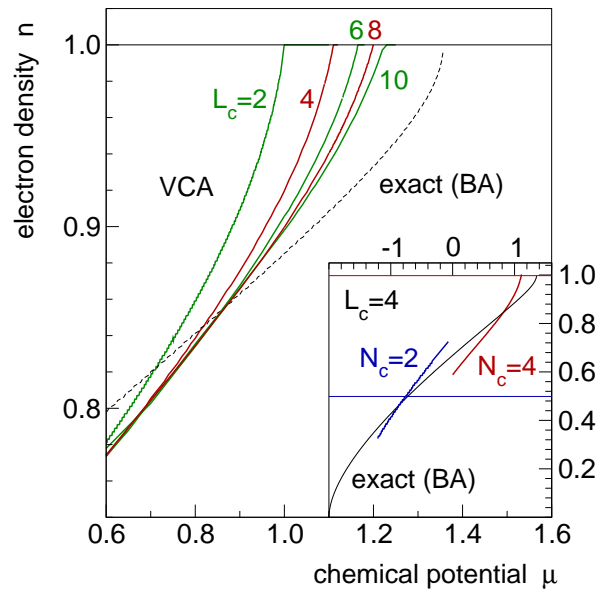}
\caption{(Color online) 
Electron density (filling $n$) as a function of the chemical potential close 
to half-filling for $U=4$ as obtained from the VCA with reference system A of 
Fig.\ \ref{fig:refsys2} for different cluster sizes $L_c$.
In all cases the cluster ground state is in the $N_c=L_c$ subspace. 
Inset: $n(\mu)$ for $L_c=4$ displayed for the entire $\mu$ range and with 
cluster ground state in the $N_c=L_c$ and in the $N_c=L_c/2$ subspace.
}
\label{fig:conv}
\end{figure}

The situation changes completely when using the variational cluster approximation. 
The most simple reference system consists of isolated clusters where only an
overall shift of the on-site energies is taken as a variational parameter to 
guarantee thermodynamical consistency, see Fig.\ \ref{fig:refsys2}, A.
The corresponding results for clusters of different size $L_c$ are shown in 
Fig.\ \ref{fig:conv}.

Apart from thermodynamical consistency, the most important difference as compared 
to the CPT, consists in the fact that a continuous $\mu$ dependence is found in 
the entire range from $\mu = -2$ (empty band) to $\mu=\mu_c$ (half-filling). 
This has explicitly been checked for clusters with $L_c=2,4,6,8,10$.
It turns out that the optimal value for the shift of the on-site energies (partly) 
compensates for the deviation of the chemical potential from its particle-hole symmetric
value $\mu = U/2$ such that (due to the large Mott-Hubbard gap) the cluster ground
state is always found in the subspace where the cluster itself is half-filled 
($N_c=L_c$). 
The cluster ground state, the optimal self-energy and eventually the filling $n$ 
thereby continuously depend on $\mu$.
This mechanism also works down to $\mu=-2$.

Fig.\ \ref{fig:conv} displays the critical regime close to half-filling only. 
We find that the critical value of the chemical potential $\mu_c$ where the transition
from the metal to the Mott-insulating state takes place, increases with increasing
cluster size $L_c$ and converges to the exact result.
The figure also shows, however, that this convergence is rather slow.
Furthermore, it remains unclear whether or not the compressibility divergence can be
recovered in the limit $L_c\to\infty$: 
Note that the slope of $n(\mu)$ (i.e.\ $\kappa$) for $n \to 1$ appears to decrease
with increasing $L_c$.
This might be explained by the argument that, even for $L_c \to \infty$ and even with
optimized cluster one-particle parameters, a trial self-energy taken from the $N_c=L_c$ 
subspace cannot describe the physics of a metallic state entirely correct.

We conclude that the continuous dependence on $\mu$ is achieved at the cost of fixing
the cluster ground state to half-filling.
This problem becomes more and more severe with decreasing filling.
The inset of Fig.\ \ref{fig:conv} shows a calculation for $L_c=4$ illustrating 
this issue.
For the calculation with $N_c=L_c=4$, the filling is reasonably close to the exact 
filling in the vicinity of half-filling only.
In the vicinity of quarter filling, however, a much better result is obtained with a
VCA calculation starting from a cluster ground state with $N_c=2=L_c/2$ (quarter-filled 
cluster).
This is physically plausible.

\subsection{Bath degrees of freedom}

As has been seen in the discussion of the results for half-filling, bath sites can considerably
help to improve a cluster approximation.
This is all the more important in the case of a metallic system off half-filling since bath sites 
also serve as charge reservoirs.
Varying the chemical potential or another physical model parameter, the electron density on the
correlated sites can vary smoothly by a charge flow from and to the uncorrelated bath sites in 
the reference system. 

Opposed to (cellular) DMFT, the filling in the original model $n$, as calculated from the approximate
VCA lattice Green's function, is generally different from the density at the correlated sites in 
the (cluster) reference system $n'$.
By rule of thumb, however, the deviations are small, i.e.\ $n \approx n'$.
This implies that all fillings from $n=0$ up to $n=1$ (half-filling) can be realized by using a
strictly {\em half-filled} reference system provided that the cluster includes (at least) one 
bath site per correlated site ($n_s=2$).
Consider, for example, a cluster with $L_c=2$ correlated and $n_s-1=1$ bath site per correlated 
site. 
In total the cluster then consists of $L'=4$ sites.
For particle-hole symmetric parameters, the cluster ground state lies in the subspace with 
$N'_{\rm tot} = 4$ electrons, and symmetry arguments imply an electron density $n'=1$ on the 
correlated and $n'_{\rm bath}=1$ on the bath sites.
This corresponds to half-filling, $n=1$, for the original model.
For $n<1$ we will find $n' \approx n <1$ and $n'_{\rm bath} > 1$ such that $n' + n'_{\rm bath} = 2$,
i.e.\ a half-filled cluster ground state.
In the limit $n\to 0$, the $N'_{\rm tot} = 4$ electrons will mostly be located on the bath sites, 
i.e.\ $n' \to 0$ and $n'_{\rm bath} \to 2$.
Analogous arguments hold for fillings above half-filling.
This mechanism promises continuous dependencies on the chemical potential with a cluster ground 
state staying in the $N'_{\rm tot} = L'$ subspace, while the physical properties are governed by 
the density on the correlated sites $n' \approx n$, which varies smoothly with $\mu$.

\begin{figure}
\centering
\includegraphics[width=0.85\columnwidth]{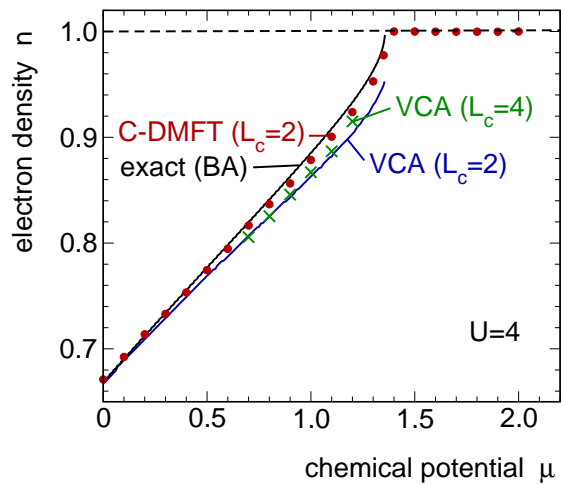}
\caption{(Color online) 
Filling $n$ as a function of $\mu$ close to half-filling for $U=4$ as obtained from 
the VCA with reference system B of Fig.\ \ref{fig:refsys2} for $L_c=2$ and $L_c=4$.
Exact and C-DMFT result from Ref.\ \onlinecite{CCK+04} are shown for comparison.
}
\label{fig:bath}
\end{figure}

We have not been able to find a stationary point by using two bath sites per correlated site
($n_s=3$).
Remembering the discussion of the half-filled case, this does not appear to be uncommon, since 
a reference system with an even number of bath sites per correlated site is expected to give a 
good description of the Mott insulator but not of the metal.
For the calculations we therefore concentrate on reference systems with $n_s=2$. 
For simplicity, we attach one bath site to any of the correlated sites and assume the hybridization
$V$ to be the same for all sites.
Additional variational parameters are $\varepsilon$ and $\varepsilon'$, the one-particle energies
of the correlated and of the bath sites which are assumed to be constant again. 
The reference system is displayed in Fig.\ \ref{fig:refsys2}, B. 

Numerical results for clusters with $L_c=2$ and $L_c=4$ correlated sites are shown in 
Fig.\ \ref{fig:bath}.
Irrespective of the cluster size, there is an excellent agreement with the exact result
for fillings lower than $n \sim 0.75$.
But also for higher fillings the VCA results with bath sites  are convincing and represent
a considerable improvement as compared to the CPT results (Fig.\ \ref{fig:cptndep}) but also
compared to the VCA results without bath sites (Fig.\ \ref{fig:conv}).
The critical chemical potential $\mu_c$ for the transition to the Mott insulator is somewhat
overestimated but the error is much smaller than the underestimation of $\mu_c$ within VCA
without bath sites.
More important, however, it appears that the approach correctly predicts the divergence of
the compressibility.
Unfortunately, it has turned out to be extremely difficult numerically to follow up the 
stationary point as a function of $\mu$ in the region very close to half-filling.
The C-DMFT results of Capone et al.\ \cite{CCK+04} which are also shown in the figure, are
slightly closer to the exact $n(\mu)$.
Note, however, that this has been achieved and crucially depends on a special (but physically 
motivated) choice for the distance measure which emphasizes the low Matsubara frequencies.
In contrast, our approximation is free from any adjustable parameter. 

\begin{figure}
\centering
\includegraphics[width=0.8\columnwidth]{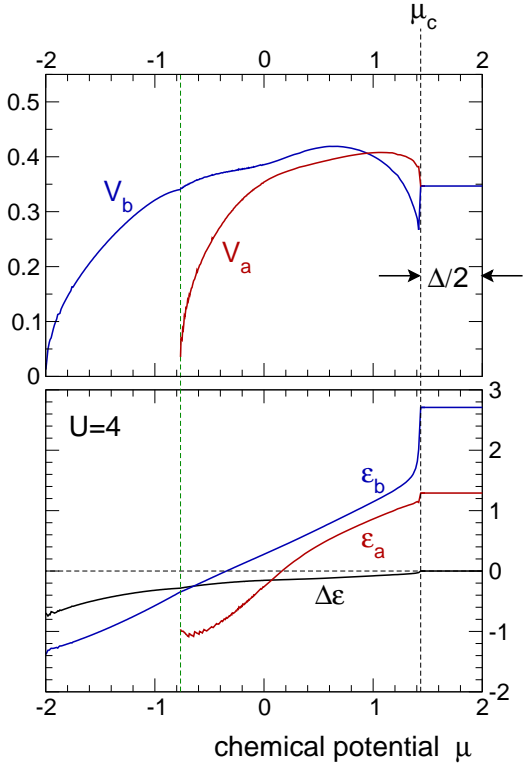}
\caption{(Color online) 
Chemical-potential dependence of the optimal one-particle parameters for reference
system C of Fig.\ \ref{fig:refsys2} at $U=4$.
$\Delta \epsilon$ is an overall shift of all one-particle energies ($\varepsilon$, 
$\varepsilon_a$ and $\varepsilon_b$).
}
\label{fig:babfig}
\end{figure}

Note that $L_c$ and the number of bath sites per correlated site (i.e.\ $n_s-1$) do not fully
specify the reference system. 
Different systems can be generated by the different ways in which bath sites are coupled to 
the correlated ones.
Reference system C in Fig.\ \ref{fig:refsys2}, for example, is characterized by $L_c=2$ and 
$n_s=2$ but spans (when independently varying all one-particle parameters) a space of trial 
self-energies which differs from the one spanned by reference system B (with $L_c=2$).
In the limit $n_s\to \infty$, i.e.\ for the case of cellular DMFT, the different ways of 
coupling the baths to the correlated sites do not matter as they can be mapped onto each 
other by unitary transformations and therefore span the same space of trial self-energies.
This is different, however, for small $n_s$. 
As is demonstrated in the following, reference systems B and C yield very similar
results for the metallic phase while C gives a much better description of the Mott 
insulator.

Any bath site in reference system C is connected via hybridizations $V_1$ and $V_2$ to 
both correlated sites.
Requiring the self-energy to be symmetric with respect to an interchange of the two 
correlated sites, implies that the modulus of the two hybridization parameters must 
be the same, i.e.\ $V_1 = \pm V_2$ (if the hybridization is assumed to be real). 
We consider two bath sites, one with $V_1 = - V_2 \equiv V_a$ and another one with
$V_1 = V_2 \equiv V_b$.
This is the only choice left if the reference system is required to respect particle-hole 
symmetry for $\mu = U/2$. 
Consequently, there are five independent variational parameters in total, $V_a$ and $V_b$, 
the bath on-site energies $\varepsilon_a$ and $\varepsilon_b$ and a general shift of all on-site
energies (including the correlated sites) $\Delta \varepsilon$.

\begin{figure}[t]
\centering
\includegraphics[width=0.95\columnwidth]{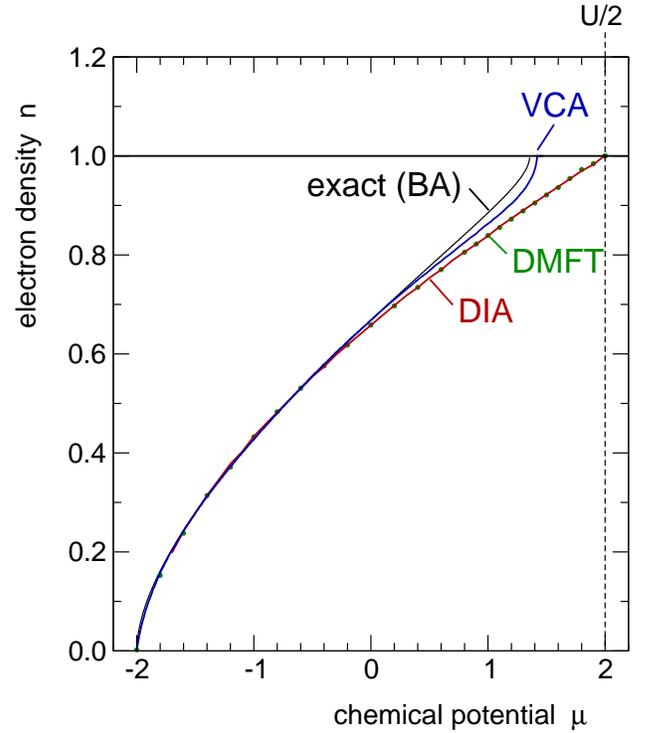}
\caption{(Color online) 
Filling versus chemical potential for $U=4$. 
VCA result using reference system C of Fig.\ \ref{fig:refsys2} in comparison with 
the exact result (Bethe ansatz), with DMFT (from Ref.\ \onlinecite{CCK+04}) and with
the two-site dynamical impurity approximation (DIA).
}
\label{fig:babnmu}
\end{figure}

Fig.\ \ref{fig:babfig} shows the optimal values of these parameters as functions of the
chemical potential.
At half-filling in the Mott insulator, i.e.\ for $\mu > \mu_c$, the parameters are $\mu$ 
independent. 
The overall shift of the on-site energies $\Delta \varepsilon$ vanishes, $V_a = V_b$, and 
$\varepsilon_a + \varepsilon_b = U$.
For the particle-hole symmetric point at $\mu = U/2$, these restrictions are enforced by 
symmetry and respected by the approximation. 
The parameters evolve continuously with $\mu$ except for $\mu = \mu_c \approx 1.42$, 
i.e.\ at the metal-insulator transition.
With decreasing $\mu$, the bath on-site energies decrease.
This results in an increasing
bath-site density $n'$ and thus in a decreasing filling $n$ as described above.
In the entire $\mu$ range, the reference system remains strictly half-filled.
At $\mu \approx -0.76$ the hybridization of bath site ``a'' vanishes, i.e.\ this site
decouples from the system. 
Since $\varepsilon_a + \Delta \varepsilon < \mu$ the site is completely occupied.
For lower $\mu$ the remaining reference system therefore consists of the two correlated 
and merely a single uncorrelated bath site to be filled with two electrons in total. 
As $\mu\to-2$ also the second bath site ``b'' decouples ($V_b \to 0$) taking both electrons
left such that the filling $n \to 0$. 
The bath sites perfectly do their job as charge reservoirs.

The resulting filling as a function of $\mu$ is displayed in Fig.\ \ref{fig:babnmu}.
The agreement with the exact Bethe ansatz result is excellent for low and intermediate 
fillings.
A slight deviation is found for fillings $n>0.8$.
Nevertheless, the qualitative trend is predicted correctly including the divergence
of the compressibility.
As compared to the results of reference system B, there is hardly any difference on the
scale of the figure.

The critical chemical potential turns out to be $\mu_c \approx 1.42$ within the VCA. 
This should be compared with the VCA result (using the same reference system) 
for the Mott-insulating gap $\Delta=1.128$ which is read off from the LDOS at half-filling.
Via
\begin{equation}
  \Delta = 2 (U/2 - \mu_c)
\end{equation}
this implies
a critical chemical potential of $\mu_c=1.436$. 
The small difference in the results for $\mu_c$ is most probably caused by numerical 
problems to locate the stationary point for the critical regime in the metallic phase
near the transition.
The predicted gap is close to the gap found with reference system K in Fig.\ \ref{fig:refsys}
for $L_c=2$ which implies that, opposed to reference system B (Fig.\ \ref{fig:refsys2}), not 
only the metallic phase 
in the entire filling range but also the Mott insulator is well described.
Considering the small size of the cluster this is actually a surprisingly good and very 
satisfying result.

We also note that the VCA calculation fully respects Luttinger's sum rule \cite{LW60}
in the entire filling range. 
While for a translationally invariant system the sum rule states that the filling equals 
the range in reciprocal space enclosed by the Fermi points, a generalized form of the sum 
rule (given in Ref.\ \onlinecite{OBP07}) must be considered here to account for the 
artificially reduced symmetry resulting from the cluster approximation. 
The main idea is that the sum rule can be derived from the equation 
$\lim_{T\to 0} \Tr ( \ff \Sigma \partial \ff G / \partial (i\omega_n) ) = 0$.
This can be tested even if the approximate self-energy and the (via Dyson's equation) 
related Green's function do not respect the translational symmetries of the underlying 
lattice and even if the system under consideration is finite.
We find that the sum rule already holds for the reference system C itself. 
As has been discussed recently, \cite{OBP07} due to the conserving nature of the VCA, 
the validity of the sum rule for the reference system is then transferred to the approximate 
quantities for the lattice problem.

Fig.\ \ref{fig:babnmu} also includes the prediction of single-site DMFT.
While close to half-filling the mean-field theory fails completely, hardly any difference to 
the exact result can be detected for fillings lower than $n \approx 0.6$.
It is worth mentioning that the results of full DMFT are {\em quantitatively} recovered 
by the most simple dynamical impurity approximation (DIA) within the framework of SFT,
namely by using a reference system consisting of the single correlated and a single bath
site only ($L_c=1$, $n_s=2$).
Both $n(\mu)$ curves, taken from DMFT and from the DIA, are identical on the scale of the figure.

\section{Conclusions}
\label{sec:sum}

Cluster mean-field theories that are based on full diagonalization or on the Lanczos technique
to treat the effective cluster problem, exhibit a number of advantages:
They directly work at zero temperature, they are flexible and can equally well treat arbitrary 
geometries, and they provide the numerically exact solution of the cluster within 
a comparatively short CPU time.
These advantages are achieved at the cost of a strongly limited cluster size (of the order of 
10 sites) dictated by the exponential dependence of the cluster Hilbert-space dimension on 
system size. 
It is therefore of highest importance to make use of the cluster degrees of freedom in the 
best possible way. 
This can be accomplished with the self-energy-functional theory.
The SFT allows to fix the cluster one-particle parameters with the help of a physical variational 
principle that is constructed for the optimization of the one-particle excitation properties.

The actual choice of the reference system, however, is not prescribed by the approach itself, 
i.e.\ different cluster topologies and thus different cluster approximations are conceivable.
This is the motivation for the present study.
With the focus on the interaction- and filling-dependent Mott metal-insulator transition, as
a prime example of a correlation effect, different reference systems have been tested against 
each other and against exact results available for the one-dimensional case.
Note that a one-dimensional lattice model actually represents the most difficult test case 
for a cluster approximation.

In the following we recapitulate the main results of our study.
First of all, the Mott insulating state of the model at the particle-hole symmetric point is 
well described by a cluster approximation using a few correlated sites only.
In particular, the ground-state energy can be determined precisely using finite-size scaling.
Already the most simple implementation of the variational-cluster approach (VCA) including 
merely an overall optimization of the intra-cluster hopping, clearly improves on the ``direct'' 
cluster approximation.
However, as compared to plain cluster-perturbation theory without any parameter optimization 
at all (but combined with the expression for the SFT grand potential), the VCA yields a marginally 
improved ground-state energy only.

Similarly, the independent optimization of several or even of all intra-cluster nearest-neighbor 
hopping parameters leads to a gain in binding energy but this is small compared to the gain 
obtained by increasing the cluster size by two more sites.
With increasing distance to the chain edges, the optimized hopping quickly converges to the 
``physical'' value, and bulk properties are already found for sites at a distance of more than 
two nearest-neighbor units from the cluster edge.
Hopping parameters vanishing in the original model can acquire a non-zero but small value in 
the optimized reference system.
Hopping parameters linking the edges of a cluster are found to vanish, i.e.\ the VCA prefers 
open boundary conditions.
The same holds for hopping parameters which would imply a breaking of particle-hole symmetry:
A next-nearest-neighbor hopping, for example, turns out to be zero at the stationary point.

The VCA correctly predicts a Mott insulating state for any $U>0$. 
It also gives a reasonable estimate for the size of the single-particle insulating gap.
This estimate improves with increasing interaction strength.
While for strong and intermediate coupling the optimal results (for the ground-state 
energy as well as for the gap) are obtained for reference systems with parameters close to the 
original model, strongly deviating parameters (more than 100\%) are favorable in the weak-coupling 
regime.
Eventually, for $U\to 0$ the VCA (and presumably any real-space cluster approach) fails to 
describe the low-energy physics of the Mott {\em transition}.
Remembering the cluster mean-field nature of the approximation, this had to be expected.
It is obvious that critical behavior cannot be accessed while, on the other hand, it is 
satisfying that physical properties on a higher energy scale are accurately captured 
with rather small clusters only. 

The additional consideration of bath sites in the reference system always yields an improved 
ground-state energy.
It has turned out that bath sites tend to decouple from a cluster reference system at the
correlated sites in the cluster center while they tightly couple to the system at the chain
edges.
As concerns the ground-state energy, however, this hardly speeds up the convergence to the 
exact result with increasing cluster size.
Going to the next larger cluster is always found to be more effective.

On the contrary, bath sites decisively influence the description of the one-particle excitation 
spectrum.
Cluster reference systems with an even number of additional bath sites coupled to a correlated site 
(at the cluster edge) give a much better result for the gap than clusters without bath sites but
more correlated sites.
This demonstrates the importance of local correlations for the one-particle spectrum. 
An odd number of bath sites per correlated site overemphasizes the mean-field character and 
thereby fails to predict the gap accurately. 

The overall features of the local density of states (insulating gap, moments, etc.) are 
addressed when looking at the local one-particle Green's function on the imaginary frequency 
axis.
Using essentially exact dynamical DMRG results as a benchmark, we find our VCA results to be 
fully competitive with cellular DMFT calculations for the same $L_c$.
While C-DMFT combined with a stochastic technique as a cluster solver is much more time-consuming,
C-DMFT ($L_c=2$, $n_s=3$) combined with Lanczos is less reliable as the VCA ($L_c=2$, $n_s=3$). 
The latter gives a significantly better result for the insulating gap which, apart from the choice
of the reference system, is unbiased and free of parameter fitting.
This shows that the thermodynamically consistent determination of the cluster parameters is 
worth the effort.
Predicting the detailed shape of the local density of states (for real frequencies) is beyond 
the capabilities of a cluster approach based on the Lanczos method: Finite-size effects
are clearly present.

As concerns characteristic quantities of the Mott insulator at half-filling, one can state 
that the convergence to the exact result is mainly determined by the number of correlated 
sites.
The additional inclusion of a continuum of bath degrees of freedom or of a large number of
bath sites appears to be unnecessary.
A {\em few} bath sites, however, can strongly improve the results.
For the description of the metallic phase off half-filling and of the filling-dependent Mott 
transition this is even more correct:

Without the inclusion of bath degrees of freedom, the VCA with optimization of an overall 
on-site energy shift predicts a smooth dependence of the filling $n$ on the chemical potential 
$\mu$. 
While this represents a clear advantage as compared to the ``direct'' cluster approach and to
cluster-perturbation theory, the main problem consists in the fact that the {\em cluster}
particle number remains constant.
Consequently, starting from a half-filled cluster, results are less and less reliable for increasing 
hole concentration $1-n$.

On the other hand, for reference systems with at least a single bath site per correlated site 
($n_s \ge 2$) there is a mechanism which solves the problem:
With the {\em total} cluster particle number always being equal to the total number of cluster 
sites, $N_{\rm tot} = L'$, the particle density at the {\em correlated} sites and, related to 
that, also the filling of the original system may vary continuously in its entire range.

For clusters with $L_c$ correlated sites and $n_s-1=1$ additional bath site attached to each, 
this mechanism has turned out to work over a wide range of fillings. 
Excellent agreement with the exact $n(\mu)$ curve from the Bethe ansatz is found for fillings
lower than $n\approx 0.75$. 
For fillings close to half-filling in the critical regime, the VCA still gives a qualitatively
satisfying result, comparable to C-DMFT calculations, but slightly overestimates the critical 
chemical potential. 
While accessing the critical regime for $n\to 1$ poses difficulties, the compressibility 
divergence is clearly visible.

A smooth $\mu$ dependence over the complete filling range as well as a good description of 
the Mott insulator can be achieved when using 
(for $L_c=2$ and $n_s=2$) a somewhat different cluster topology where a bath site couples 
to {\em both} correlated sites.
We find that both the exact result as well as the best (DMFT) mean-field result are almost
quantitatively recovered with small ($L_c=2$, $L_c=1$) reference systems including the 
minimum number ($n_s=2$) of local degrees of freedom. 

The one-dimensional Hubbard model considered in this study has merely served for benchmarking 
purposes.
Eventually, our main interest is focused on the physics of strongly correlated electrons on 
two- or higher-dimensional lattices. 
Doped Mott insulators in two dimensions, however, constitute notoriously difficult many-body 
problems which require, as a prerequisite, thorough studies of systems which are more controlled
in order to avoid artifacts or misinterpretations.

\acknowledgments
Instructive discussions with M. Aichhorn, C. Dahnken and H. Hafermann are gratefully 
acknowledged.
The work is supported by the Deutsche Forschungsgemeinschaft within the 
Forschergruppe FOR~538.

\end{document}